\documentclass[twocolumn,showpacs,preprintnumbers,amssymb]{revtex4}

\usepackage{amsmath2000}
\usepackage{graphicx}
\usepackage{dcolumn}
\usepackage{bm}

\begin{document}

\preprint{}

\title{Transverse self-fields within an electron bunch moving in an arc of a circle}

\author{Gianluca Geloni}
\email{g.a.geloni@tue.nl}

\author{Jan Botman}
\author{Jom Luiten}
\author{Marnix van der Wiel}

\affiliation{%
Department of Applied Physics, Technische Universiteit Eindhoven,
\\
 P.O. Box 513, 5600MB Eindhoven, The Netherlands
}%

\author{Martin Dohlus}
\author{Evgeni Saldin}
\author{Evgeni Schneidmiller}
\affiliation{Deutsches Elektronen-Synchrotron DESY, \\
Notkestrasse 85, 22607 Hamburg, Germany
}%

\author{Mikhail Yurkov}
\affiliation{Particle Physics Laboratory (LSVE), Joint Institute for Nuclear Research, \\
141980 Dubna, Moscow Region, Russia}

\begin{abstract}
As a consequence of motions driven by external forces,
self-fields (which are different from the static case) originate
within an electron bunch. In the case of magnetic external forces
acting on an ultrarelativistic beam, the longitudinal
self-interactions are responsible for CSR (Coherent Synchrotron
Radiation)-related phenomena, which have been studied
extensively. On the other hand, transverse self-interactions are
present too. At the time being, existing theoretical analysis of
transverse self-forces deal with the case of a bunch moving along
a circular orbit only, without considering the situation of a
bending magnet with a finite length. In this paper we propose an
electrodynamical analysis of transverse self-fields which
originate, at the position of a test particle, from an
ultrarelativistic electron bunch moving in an arc of a circle.
The problem will be first addressed within a two-particle system.
We then extend our consideration to a line bunch with a stepped
density distribution, a situation which can be easily generalized
to the case of an arbitrary density distribution. Our approach
turns out to be also useful in order to get a better insight in
the physics involved in the case of simple circular motion and in
order to address the well known issue of the partial compensation
of transverse self-force.

\end{abstract}


\begin{widetext}
\thispagestyle{empty}
\begin{large}
\textbf{DEUTSCHES ELEKTRONEN-SYNCHROTRON}\\
\end{large}

DESY 02-048

May 2002

\begin{eqnarray}
\nonumber &&\cr \nonumber && \cr \nonumber &&\cr
\end{eqnarray}

\begin{center}
\begin{Large}
\textbf{Transverse self-fields within an electron bunch moving
\\in an arc of a circle}
\end{Large}
\begin{eqnarray}
\nonumber &&\cr \nonumber && \cr
\end{eqnarray}

\begin{large}
Gianluca Geloni, Jan Botman, Jom Luiten and Marnix van der Wiel
\end{large}

\textsl{\\Department of Applied Physics, Technische Universiteit
Eindhoven, \\P.O. Box 513, 5600MB Eindhoven, The Netherlands}
\begin{eqnarray}
\nonumber
\end{eqnarray}
\begin{large}
Martin Dohlus, Evgeni Saldin and Evgeni Schneidmiller
\end{large}

\textsl{\\Deutsches Elektronen-Synchrotron DESY, \\Notkestrasse
85, 22607 Hamburg, Germany}
\begin{eqnarray}
\nonumber
\end{eqnarray}
\begin{large}
Mikhail Yurkov
\end{large}

\textsl{\\Particle Physics Laboratory (LSVE), Joint Institute for
Nuclear Research, \\141980 Dubna, Moscow Region, Russia}

\newpage

\end{center}
\end{widetext}

\maketitle

\section{\label{sec:intro}INTRODUCTION}

When an electron bunch undergoes a motion under the influence of
external forces, the particles within the bunch become sources of
self-fields which are different from the static case and which
obey the usual Lienard-Wiechert expressions.

These self-fields feed back on the particle dynamics, which often
makes the description of the evolution of the system a problem
without easy solution.

Many electrodynamical systems fit in the latter description. In
all these systems one can recognize two separate aspects in the
evolution issue: a dynamical one, which is governed by the
equation of motion, and an electrodynamical one, which is taken
care of by Maxwell equations. The solution to the evolution
problem is obtained when one is able to solve simultaneously both
equations: this is, for example, what self-consistent computer
codes strive for. An example of such self-consistent, numerical
solutions is given (see \cite{ROTH}) by the program TRAFIC$^{4}$
(which will be employed in this paper to provide cross-checks with
analytical results). Although TRAFIC$^{4}$ and other codes are
able to provide quite reliable results for a number of evolution
problems, a separate study of their dynamical and electrodynamical
aspects, which can be achieved by means of theoretical analysis,
is necessary in order to get a good understanding of the physics
involved,  and to provide benchmark and cross-checks for
simulations (and, at an earlier stage, to build correct
simulations too).

Moreover, it often happens that this kind of theoretical analysis
can be directly applied in order to get good practical solutions
for the problem at least in a narrow region of the parameters
which specify the system setup. The results can, then, be used for
quick estimations of the magnitude of the effects under
investigation.

Let us restrict our attention to the case of self-fields within an
ultrarelativistic electron bunch driven by external forces of
magnetic nature.

The self-interaction in the longitudinal direction (parallel, at
any time, to the velocity vector by definition) is responsible for
the energy exchange between the system and the acceleration field
and for all CSR (Coherent Synchrotron Radiation)-related
phenomena, which have been studied extensively elsewhere (see,
among the others, \cite{ROTH}... \cite{BORL}).

These investigations are important in view of the need for very
high-peak current, low emittance beams for self-amplified
spontaneous emission (SASE)-free-electron-lasers operating in the
x-ray region (see, among the others, \cite{XRAY}). Similar beams
are also being considered for production of femtosecond radiation
pulses by simpler schemes based on Cherenkov and Transition
Radiation \cite{WALTER}: production and utilization of such kind
of beams may prove difficult due to self-field collective effects.
Indeed, these effects may spoil the required high brightness of
the electron beam, which is a matter of major concern among people
involved this kind of physics research.

The study of self-forces in the transverse direction is important
for the same reasons. They were first addressed, in the case of a
circular motion, and from an electrodynamical viewpoint, in
\cite{TALM}. Further analysis (\cite{RUI2}, \cite{LEE}...
\cite{STUP}) consider, again, the case of circular motion both
from an electrodynamical and a dynamical viewpoint and in the
approximation of a rigid bunch: at the time being, existing
theoretical analysis of transverse self-forces deal with the case
of circular orbit only, without considering transient collective
phenomena.

In this paper we propose a fully electrodynamical analysis of
transverse self-fields originating, at the position of a test
particle, from an ultrarelativistic electron bunch moving in an
arc of a circle, thus treating for the first time, besides the
basic situation of circular motion, also the case of transient
between a straight line and a hard-edge magnet (and, vice versa,
from a bend to a straight line).

Consistently with the choice to analyze the electrodynamical
aspect of the problem only, a zero energy-spread will be
understood when considering the evolution of an electron bunch. As
underlined before, although the results obtained can be directly
applied, from a practical viewpoint, only in the case in which
the zero energy-spread hypothesis is verified \textit{a
posteriori}, the outcomes of this paper are important to get a
good insight if the physics of the problem, and to provide
benchmark and cross-checks for simulations.

Firstly, a two-particle model is adopted in order to study the
transverse force produced by a single particle, and then, by
summing up all the contributions from different electrons, the
case of a line-bunch model characterized by a rectangular density
distribution is analyzed: this can, in fact, be easily generalized
to the case of an arbitrary density distribution.

Besides providing results in the case of a finite hard-edge
bending magnet, our approach turns also useful in order to get a
better insight of the physics involved in the case of simple
circular motion.

Moreover, our results give us a better understanding (although
one has to take into account, here, dynamical aspects as well as
electrodynamical ones) of the partial compensation between
transverse self-force and gradient of the potential energy
deviation from the nominal value.


We will discuss the limits of applicability of our model, with
respect to the transverse beam size, $h$, in Section
\ref{sec:steadybunch}). Within such applicability limits, we find
results which are independent of the bunch transverse dimension.

The paper is organized as follows. The transverse interaction
between two electrons moving on a circle, together with a simple
dynamical interpretation is treated in Section
\ref{sec:steadytwo}. In Section \ref{sec:steadybunch}, we deal
with a stepped-profile electron bunch interacting with a test
particle again on a circle, and discuss also the applicability
region of the line model. Transient behavior (from straight to
circular path and vice versa) for the transverse self-forces
between two particles are then studied in Section
\ref{sec:transientwo}. Results for the transient of a
stepped-profile bunch are given in Section
\ref{sec:transientbunch}, where a treatment for the case of a more
generic bunch density is also proposed. A regularization
technique for cancelling the singularity in the expression for
the transverse force (which always arises in the limit of a
near-zero distance between test particle and sources) is then
applied. In Section \ref{sec:cancel} we will deal with the
well-known issue of the partial compensation of transverse
self-force.
Finally, in Section \ref{sec:concl}, we come to a summary of the
obtained results and to conclusions.

\section{\label{sec:steadytwo}TRANSVERSE INTERACTION BETWEEN TWO ELECTRONS MOVING IN A CIRCLE}

Let us begin our study considering the steady case of  two
electrons moving on a circle of radius $R$. The electro-magnetic
force which one of the two particles (designated with "T", i.e.
the test particle) feels, due to the interaction with the other
one (designated with "S", i.e. the source particle), is given by

\begin{equation}
{\bm F}({\bm {r_{\mathrm{T}}}},t) = e{\bm E}({\bm
{r_{\mathrm{T}}}},t) + ec{\bm{\beta_{\mathrm{T}}}} \times {\bm
B}({\bm {r_{\mathrm{T}}}},t) \label{lorentz},
\end{equation}
where ${\bm {r_{\mathrm{T}}}}$ is the position of the test
particle, $e$ is the electron charge with its own (negative) sign,
$\bm{\beta_{\mathrm{T}}}$ is the velocity of the test particle
normalized to the speed of light, $c$, while $\bm{E}({\bm
{r_{\mathrm{T}}}},t)$ and $\bm{B}({\bm {r_{\mathrm{T}}}},t)$ are,
respectively, the electric and the magnetic field generated at a
given time $t$ by the source particle S, at the position of the
test particle T, namely

\begin{eqnarray} {\bm E}({\bm {r_{\mathrm{T}}}},t)={e\over {4 \pi
\varepsilon_\mathrm{0}}}\Biggl\{{1 \over \gamma_\mathrm{S}^{2}}
{{\bf {\hat{n}}} - {\bm {\beta_\mathrm{S}}}
\over{R_\mathrm{ST}^{2} \left({1-{\bf {\hat n}} \cdot {\bm
{\beta_\mathrm{S}}}}\right)^{3}}} \nonumber\\&\cr + {1\over c}
{{\bf{\hat n}} \times \left[{\left({{\bf{\hat n}}-{\bm
{\beta_\mathrm{S}}}}\right)\times {\dot{\bm
\beta}_{\bm{\mathrm{S}}}}}\right] \over{R_\mathrm{ST} \left({1-
{\bf{\hat n}} \cdot {\bm {\beta_\mathrm{S}}}}\right)^{3}}}
\Biggr\}~ && \label{Efield}
\end{eqnarray}
and

\begin{equation}
{\bm B}({\bm {r_{\mathrm{T}}}}, t)={1\over c} {\bf{\hat n}}
\times {\bm E}({\bm {r_{\mathrm{T}}}},t)~. \label{Mfield}
\end{equation}
Here ${\bm{\beta_\mathrm{S}}}$ and ${\dot{\bm
\beta}_{\bm{\mathrm{S}}}}$ are, respectively, the dimensionless
velocity and its time derivative at the retarded time $t'$,
$R_\mathrm{ST}$ is the distance between the retarded position of
the source particle and the present position of the test electron,
$\bf{\hat{n}}$ is a unit vector along the line connecting those
two points and $\gamma_\mathrm{S}$ is the usual Lorentz factor
referred to the source particle at the retarded time $t'$.

\begin{figure}
\includegraphics*[width=84mm]{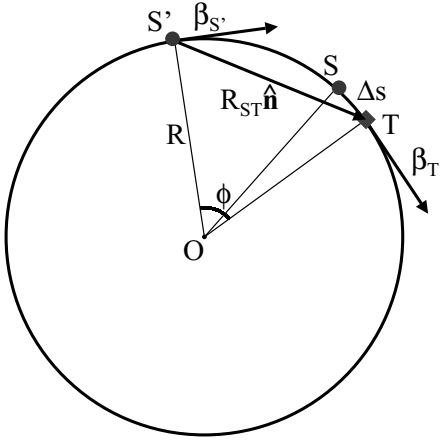}
\caption{\label{FIG1} Geometry for the two-particle system in the
steady state situation, with the test particle ahead of the
source. Here T is the present position of the test particle, S is
the present position of the source, while S' indicates the
retarded position of the source.}
\end{figure}

The transverse direction (on the orbital plane) is, by
definition, orthogonal to $\bm{\beta_\mathrm{T}}$. The transverse
component of Eq. (\ref{lorentz}) can be written as the sum of
contributions from the velocity ("C", Coulomb) and the
acceleration ("R", Radiation) fields, namely

\begin{equation}
{\bm {F_\bot}}({\bm {{r_{\mathrm{T}}}}},t) = {\bm {F_\mathrm{\bot
C}}}({\bm {r_{\mathrm{T}}}},t)+{\bm {F_\mathrm{\bot R}}}({\bm
{r_{\mathrm{T}}}},t), \label{transvlorentz}
\end{equation}
where

\begin{equation}
{\bm F_{\bm{\mathrm{\bot C}}}}({\bm {r_{\mathrm{T}}}},t) = {e^2
\over {4 \pi \varepsilon_\mathrm{0}}} {{\bf{n_\bot}}\left(1 -
{\bm{\beta_\mathrm{S}}}\cdot{\bm{\beta_\mathrm{T}}}\right) -
{\bm{\beta_\mathrm{\bot S}}}\left(1 - {\bf{\hat{n}}}\cdot
{\bm{\beta_\mathrm{T}}}\right)\over{\gamma_\mathrm{S}^2
R_\mathrm{ST}^2
\left(1-{\bf{\hat{n}}}\cdot{\bm{\beta_\mathrm{S}}}\right)^3}}
\label{Coulomb}
\end{equation}
and
\begin{widetext}
\begin{equation}
{\bm F_{\bm{\mathrm{\bot R}}}}({\bm {r_{\mathrm{T}}}},t) = {e^2
\over {4 \pi \varepsilon_\mathrm{0}}} \Bigg[ { {\bf{n_\bot}}
\left({\bf{\hat{n}}}\cdot{{\dot{\bm{\beta}}_{\bm{\mathrm{S}}}}}\right)\left(1
- {\bm{\beta_\mathrm{S}}}\cdot{\bm{\beta_\mathrm{T}}}\right)-
{\bm{\beta_\mathrm{\perp
S}}}\left({\bf{\hat{n}}}\cdot{{\dot{\bm{\beta}}_{\bm{\mathrm{S}}}}}\right)\left(1-
{\bf{\hat{n}}}\cdot{\bm{\beta_\mathrm{T}}}\right)\over{R_\mathrm{ST}
\left(1-{\bf{\hat{n}}}\cdot{\bm{\beta_\mathrm{S}}}\right)^3}} - {
\dot{\bm{\beta}}_{\bm{\mathrm{\perp T}}}\left(1 -
{\bf{\hat{n}}}\cdot{\bm{\beta_\mathrm{T}}}\right) +
{\bf{\hat{n}_\perp}}\left({\bm{\beta_\mathrm{T}}}\cdot{{\dot{\bm{\beta}}_{\bm
{\mathrm{S}}}}}\right) \over{R_\mathrm{ST}
\left(1-{\bf{\hat{n}}}\cdot{\bm{\beta_\mathrm{S}}}\right)^2}}\Bigg]\;.
\label{Radiation}
\end{equation}
\end{widetext}
Let us first consider the case in which the test particle is in
front of the source. In this case, referring to Fig. \ref{FIG1},
we can define with $\Delta s$ the curvilinear distance between
the present position of the test and of the source particle;
$\phi$ will indicate, instead, the angular distance between the
retarded position of the source and the present position of the
test electron, and it will be designated as the retarded angle.

In the following we will assume $\beta_\mathrm{S} =
\beta_\mathrm{T} = \beta$. This hypothesis will naturally lead,
further on, to the assumption of a zero energy-spread when one
considers the evolution of an electron bunch, and, as already
discussed in Sec. \ref{sec:intro}, it is consistent with our
choice to analyze only the electrodynamical aspect of the
problem. Upon this we can write Eq. (\ref{Coulomb}) and Eq.
(\ref{Radiation}) in the following way:

\begin{equation}
{F_\mathrm{\bot C}} = {e^2 \over {4 \pi \varepsilon_\mathrm{0}}}
{1+\beta^2 - 2\beta \cos(\phi/2) \over{4 R^2 \gamma^2
\sin(\phi/2) (1-\beta\cos(\phi/2))^3}}~, \label{Coultrig}
\end{equation}
\begin{eqnarray}
{F_\mathrm{\bot R}} = {e^2 \over {4 \pi \varepsilon_\mathrm{0}}}
{\beta^2\over{2 R^2}}\Bigg[ {1\over{\sin(\phi/2) \left(1-\beta
\cos(\phi/2) \right)}} \nonumber\\&\cr -
{(1-\beta^2)\sin{\phi/2}\over{\left(1-\beta \cos(\phi/2)
\right)^3}}\Bigg]~. &&\label{Radtrig}
\end{eqnarray}
If we now assume a small retarded angle ($\phi\ll 1$), which is
justified in the ultrarelativistic case, we can expand Eq.
(\ref{Coultrig}) and Eq. (\ref{Radtrig}) to the second
non-vanishing order in $\phi$ thus obtaining

\begin{equation}
{F_\mathrm{\bot C}} \simeq {e^2 \gamma^3 \over {4 \pi
\varepsilon_\mathrm{0} R^2}} \Phi_\mathrm{C}(\hat{\phi})
\label{Coulexp}
\end{equation}
and

\begin{equation}
{F_\mathrm{\bot R}} \simeq {e^2 \gamma^3 \over {4 \pi
\varepsilon_\mathrm{0} R^2}} \Phi_\mathrm{R}(\hat{\phi})~,
\label{Radexp}
\end{equation}
where we define $\Phi_\mathrm{C}$ and $\Phi_\mathrm{R}$ as

\begin{equation}
\Phi_\mathrm{C}(\hat{\phi}) = {
\hat{\phi}^2 \over
{\hat{\phi} (1+\hat{\phi}^2/4)^3}} \label{PhiC}
\end{equation}
and

\begin{equation}
\Phi_\mathrm{R}(\hat{\phi}) = {2-\hat{\phi}^2+\hat{\phi}^4/8
\over{\hat{\phi} (1+\hat{\phi}^2/4)^3}} ~.\label{PhiR}
\end{equation}
Here and above $\hat{\phi} = \gamma \phi$. This normalization
choice, already treated in \cite{SALLONG}, is quite natural,
$1/\gamma$ being the synchrotron radiation formation-angle at the
critical wavelength. In the derivation of Eq. (\ref{Coulexp}) and
Eq. (\ref{Radexp}) (and in the following, too) we understood $\hat
\phi \gg 1/\gamma$, which is again justified by the
ultrarelativistic approximation. Moreover, it is important to
realize that the assumption $\phi \ll 1$ is by no means a
restrictive one, since it keeps open the possibility of comparing
$\phi$ with the formation angle $1/\gamma$ (note that a
deflection angle smaller or bigger than $1/\gamma$ is
characteristic of the cases, respectively, of undulator or
synchrotron radiation).

The following expression can be then trivially derived, which is
valid for the total transverse force felt by the test particle
\begin{equation}
{F_\mathrm{\bot}} \simeq {e^2 \gamma^3 \over {4 \pi
\varepsilon_\mathrm{0} R^2}} \Phi(\hat{\phi})~, \label{Totexp}
\end{equation}
where $\Phi$ is defined by

\begin{equation}
\Phi(\hat{\phi}) = {2+\hat{\phi}^4/8 \over{\hat{\phi}
(1+\hat{\phi}^2/4)^3}} ~,\label{Phi}
\end{equation}
Note that Eq. (\ref{Phi}) is completely independent of the
parameters of the system: it is then straightforward to study the
asymptotic behaviors of $\Phi$. In order to do so, just remember
that the retardation condition linking $\Delta s$ and $\phi$ is
given by (see \cite{SALLONG})

\begin{equation}
\Delta s = R\phi -2\beta R\sin{\phi\over{2}}~, \label{retsteady}
\end{equation}
or its approximated form

\begin{equation}
\Delta s = (1-\beta)R\phi + {R\phi^3\over{24}}~.
\label{retsteadyappr}
\end{equation}
It is now evident that $\Phi(\Delta \hat{s}) \rightarrow
1/(3\Delta \hat{s})$ when $\hat{\phi} \gg 1$ and $\Phi(\Delta
\hat{s}) \rightarrow 1/(\Delta \hat{s})$ when $\hat{\phi} \ll 1$,
having introduced the normalized quantity $\Delta \hat{s} =
(\gamma^3/R) \Delta s$. This normalization choice is linked to
the fact that the critical synchrotron radiation wavelength,
$R/\gamma^3$, is also the minimal characteristic distance of our
system: two particles nearer than such a distance can be
considered as a single one radiating, up to the critical
frequency, with charge $2e$ (see \cite{SALLONG}).

\begin{figure}
\includegraphics*[width=90mm]{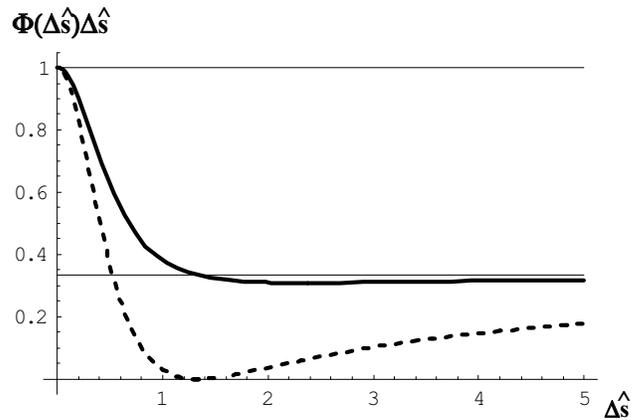}
\caption{\label{FIG2} Plot of $\Phi(\Delta \hat{s})\Delta
\hat{s}$ (solid line) and comparison with the asymptotic values,
$1$ and $1/3$. The dashed line shows the radiative contribution
$\Phi_\mathrm{R}(\Delta \hat{s})\Delta \hat{s}$ alone.}
\end{figure}

The asymptotic behavior above suggests to study the function
$\Phi(\Delta \hat{s})\Delta \hat{s}$. We plotted such a function
in Fig. \ref{FIG2} (together with the radiative contribution
alone) for values of $\Delta \hat{s}$ running from 0 to 5.

It is interesting to underline the fact that, as one can see from
Fig. \ref{FIG2}, the transverse force is always centrifugal, for
any distance between the two particles. This fact can be explained
by means of a simple relativistic argument which holds,
qualitatively, for all particle distances: in order to build the
two-particle system, i.e. to bring them together, one needs to
work against the electromagnetic field. Then, the total mass of
the system accounts for this interaction energy too, and is
therefore bigger than the simple sum of the particles masses.
Hence, also the equilibrium orbit radius must be bigger than $R$,
and a centrifugal self-force is to be expected.

It is also worthwhile to note that $\Phi_\mathrm{R}$ has the same
asymptotic behavior as $\Phi$, and that $\Phi_\mathrm{C}$ gives
important contributions to $\Phi$ only for values between the two
asymptotes (see Fig. \ref{FIG2}). This may find intuitive
explanation in the following reasoning. As is well known (see,
for example, \cite{JACK}), the velocity field of an
ultrarelativistic electron is radial, with respect to the
"virtual" position which the particle would assume if it moved
with constant velocity starting from the retarded point, but the
line forces are not isotropically distributed, and resemble more
and more the plane wave configuration as $\beta \rightarrow c$.
Therefore, the test particle is influenced by the velocity field
of the source particle only when such a field "shines right on the
test electron" (quoted from \cite{RUI1}), which does not happen
for asymptotic values of $\hat{\phi}$.

\begin{figure}
\includegraphics*[width=84mm]{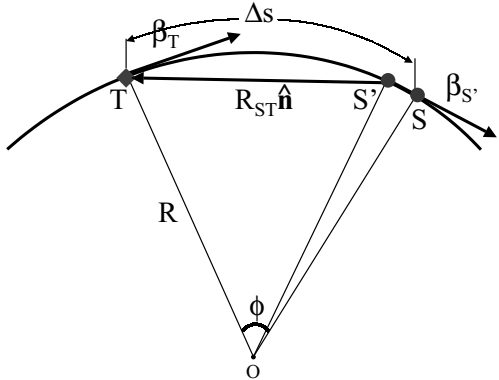}
\caption{\label{FIG3} Geometry for the two-particle system in the
steady state situation, with the source particle ahead of the test
one. Here T is the present position of the test particle, S is the
present position of the source, while S' indicates the retarded
position of the source.}
\end{figure}

Let us now analyze, in the framework of our line model, the case
in which the source particle is in front of the test particle. The
geometry is qualitatively sketched in Fig. \ref{FIG3}.

The difference with respect to the situation in which the test
electron is in front of the source is that the test electron "runs
against" the electromagnetic signal emitted by the source, while
in the other case it just "runs away" from it. Therefore the
relative velocity between the signal and the test electron is
equal to $(1+\beta)c$, instead of $(1-\beta)c$ in the other
situation. Hence the retardation condition reads

\begin{equation}
\Delta s = R\phi + 2\beta R\sin{\phi\over{2}}
\label{retcondhtexact}
\end{equation}
or, solved for $\phi$ in its approximated form,

\begin{equation}
\phi \simeq {\Delta s\over{R(1+\beta)}}~. \label{retcondht}
\end{equation}
In this situation, $\bm{\beta_\mathrm{S}}$ is almost parallel to
(and equal to) $\bm{\beta_\mathrm{T}}$ and antiparallel to
$\hat{n}$: it turns out that the only important contribution to
the transverse force is given by the second term on the right
side of Eq. (\ref{Radiation}), and it is easy to check that

\begin{equation}
{F_\mathrm{\bot}} \simeq {e^2  \over {4 \pi
\varepsilon_\mathrm{0} R \Delta s}}~.
 \label{Fperpht}
\end{equation}
It may be worthwhile to underline that the force in Eq.
(\ref{Fperpht}) is, evidently, of the same magnitude as the force
in Eq. (\ref{Totexp}).

As a note to the entire steady-state case discussion, let us
provide an interpretation for the electromagnetic transverse
forces from the viewpoint of relativistic dynamics in the limit
$\Delta s \ll R/\gamma^3$. In such a situation the two electrons
are near enough to consider them travelling with the same
velocity vector. One can, then, analyze the situation in the rest
frame of the two-particle system. By means of a Lorentz
transformation, one can find that the total momentum of the
system in the laboratory frame and in the direction of motion
(let us call it \textit{z}-direction), is given by (see
\cite{JACK}):

\begin{equation}
{P_\mathrm{z}} = \gamma\left(2m + U/c^2 - {1\over{c^2}} \int
T_\mathrm{33}~ dV\right)\beta c~,
 \label{totmom}
\end{equation}
where the integration (in Eq. (\ref{totmom}) and below) is
performed over all space. Here $m$ is the electron rest mass, $U$
is the electromagnetic interaction energy of the system and
$T_\mathrm{33}$ is the component of the electromagnetic stress
tensor corresponding to the flux of the \textit{z}-component of
the total momentum along the \textit{z} direction. It is
worthwhile to remark that we do not have to include
$T_\mathrm{13}$ and $T_\mathrm{23}$ (corresponding anyway to
momentum flux in the z-direction) in the calculation of $P_z$
because they are antisymmetric in $z$ (as well as in the other
two directions, $x$ and $y$, individuated by the tensor indexes
$1$ and $2$), thus giving a vanishing integral contribution. All
the quantities between round brackets are calculated in the system
rest frame.

On the other hand, the total energy of the system turns out to be
(see, again, \cite{JACK}):
\begin{equation}
{U_\mathrm{syst}} = \gamma\left(2mc^2 + U - \beta^2\int
T_\mathrm{33}~ dV\right),
 \label{toten}
\end{equation}
where, again, the quantities between round brackets are calculated
in the system rest frame.

It is now easy to calculate $U$ and $T_\mathrm{33}$ in the rest
frame. The electromagnetic interaction energy is given by the
work done against the field to bring the two particles together
from a situation in which they are separated by an infinite
distance:

\begin{equation}
U \simeq {e^2  \over {4 \pi \varepsilon_\mathrm{0} \gamma \Delta
s}}~.
 \label{U}
\end{equation}
In order to calculate $T_\mathrm{33}$, we can consider the
electrostatic stress tensor alone because in the rest frame, at
short distance $\gamma \Delta s \ll R/\gamma^2$, the radiative
field contributions are unimportant and, on the other hand, the
external magnetic field has zero component along the direction of
motion. Now, the electrostatic stress tensor is given by

\begin{equation}
T_\mathrm{\textit{ij}}= \varepsilon_0(E_\textit{i}E_\textit{j}-
\delta_\textit{ij}E^2/2)~, \label{stress}
\end{equation}
where \textit{i}, \textit{j} = 1... 3. For us, the only
interesting component is

\begin{equation}
T_\mathrm{33} = +E_z^2- E^2/2 ~.\label{stressint}
\end{equation}
It can be proven that

\begin{equation}
\int T_\mathrm{33} ~dV= -U ~.\label{stressequU}
\end{equation}
%

%

%

Now, the equations for the momentum and the energy of the system
in the laboratory frame read:
\begin{equation}
{P_\textit{z}} = \gamma\left(2m + 2U/c^2\right)\beta c~,
 \label{lasttotmom}
\end{equation}
and
\begin{equation}
{U_\mathrm{syst}} = \gamma\left[2mc^2 + U(1+\beta^2)\right]~.
\label{lasttoten}
\end{equation}
From the transverse component of the equation of motion for the
system one gets

\begin{equation}
F_\mathrm{\bot syst} \simeq 2 e B \beta c - 2 {e^2  \over {4 \pi
\varepsilon_\mathrm{0} R \Delta s}} \label{eqmot}
\end{equation}
which justifies Eq. (\ref{Fperpht}) and the asymptotic behavior
for short distance between the particles of Eq. (\ref{Totexp}).

The discussion above underlines the fact that, in order to get a
correct dynamical interpretation of the electrodynamical
transverse forces on the two particles, one needs to account for
the self-interaction energy and momentum flux of the system.
Nevertheless, by doing this, one can easily find that the energy
of our system (divided by $c$) and its momentum do not constitute
a four-vector anymore, as it is seen directly from Eq.
(\ref{totmom}) and Eq. (\ref{toten}).
At the beginning of the 20th century a similar problem was
encountered by people trying to build a classical model of the
electron: in that case, the solution was the introduction of the
Poincar\'e stress tensor (see, among others, \cite{JACK} or
\cite{MOLL}) , which granted stability to the system and
recovered covariance for the energy-momentum pair. As said in
\cite{JACK}: "It is not unreasonable then to include Poincar\'e
stresses in our classical models of charged particles, or at
least to remember that care must be taken in discussion of purely
electromagnetic aspects of such models". In our case we deal with
a completely electrodynamical system, so that there is no
straightforward way to introduce the analogue of Poincar\'e
stresses: nevertheless one must take care and remember that,
within our accuracy, energy and momentum are no longer forming a
four-vector. It might be worthwhile to underline the fact that,
in modern physics of particle acceleration, problems which have
been of pure academic importance for about one hundred years are
now becoming of practical importance.

\section{\label{sec:steadybunch}TRANSVERSE INTERACTION BETWEEN AN ELECTRON
AND A BUNCH MOVING IN A CIRCLE}

\begin{figure}[htb]
\includegraphics*[width=90mm]{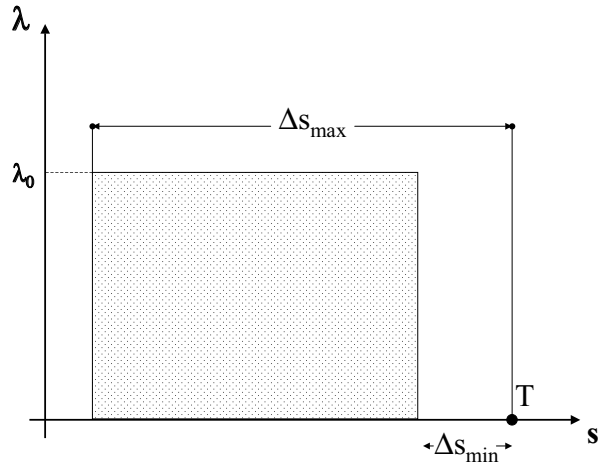}
\caption{\label{FIG4} Schematic of the test electron T and the
stepped-profile bunch; $\Delta s_\mathrm{min}$ indicates the
distance between the electron and the head of the bunch; $\Delta
s_\mathrm{max}$ is the sum of the bunch length and $\Delta
s_\mathrm{min}$}
\end{figure}

In this Section we discuss the transverse force felt by an
electron due to the interaction with a line bunch (with
rectangular density distribution) moving in a circle. The geometry
is described in Fig. \ref{FIG4}, and will be considered rigid,
i.e. fixed during the entire evolution of the system. This fact
means that we will have zero energy-spread inside our bunch and
this is in agreement with our model choice in Section
\ref{sec:steadytwo} and our program of a pure electrodynamical
analysis, without dynamical aspects.

Before beginning actual calculations, we provide here a discussion
about the applicability region of our line model. In the case of
contributions from particles behind the test particle, the region
of applicability of the line model follows straightforwardly from
the retardation condition: one can easily check that the
inclusion of a transverse dimension of the bunch, already
designated with $h$ in Section \ref{sec:intro}, adds to Eq.
(\ref{retsteadyappr}) the term of magnitude $h^2/(R\phi)$. Such a
term is negligible with respect to the others in the retardation
condition, whenever $h \ll R/\gamma^2$, which specifies the
region of applicability of our model as regards the transverse
bunch size.

The situation becomes more complicated when one considers
contributions from electrons in front of the test particle. In
fact, in the case $\Delta s < 0$ (source particle in front) and
$|\Delta s| < h$, we have a situation in which the test electron
overtakes the source  before it is reached by the electromagnetic
signal: this means that, even if the test particle is behind the
source, the present position of the test particle is, anyway, in
front of the retarded position of the source. Then, it is
possible to show that the line model constitutes a valid
description of the situation only when $|\Delta s| \gg h$. The
cases that do not verify such a condition are left for future
study.

Now, in order to actually evaluate the transverse electromagnetic
force which the bunch exercises on the test particle one has to
sum up the contributions from all the retarded sources. Since the
two-particle interaction has been calculated in Section
\ref{sec:steadytwo} as a function of the retarded angle $\phi$, it
is convenient to switch the integration variable from $\Delta s$
to $\phi$,

\begin{equation}
{d \Delta s\over{d \phi}} = R\left(1-\beta \cos{\phi\over{2}}
\right) \label{Jacobian}
\end{equation}
being the Jacobian of the transformation. Therefore, from Eq.
(\ref{Coultrig}), Eq. (\ref{Radtrig}) and Eq. (\ref{Jacobian})
one has

\begin{eqnarray}
{F_\mathrm{\bot C}^\mathrm{B}} =
\int_\mathrm{\phi_\mathrm{min}}^{\phi_\mathrm{max}} F_\mathrm{\bot
C}(\phi)~ R\left(1-\beta \cos(\phi/2) \right) d\phi =
\nonumber\\&&\cr = { e^2 \lambda_0 \over {4 \pi
\varepsilon_\mathrm{0} R}} {1\over{2 \gamma^2}} \left[
-{\beta\over{1-\beta \cos(\phi/2) }} -
\ln(\tan(\phi/4))\right]_\mathrm{\phi_\mathrm{min}}^{\phi_\mathrm{max}}&&
~ \label{intC}
\end{eqnarray}
and
\begin{eqnarray}
{F_\mathrm{\bot R}^\mathrm{B}} =
\int_\mathrm{\phi_\mathrm{min}}^{\phi_\mathrm{max}} F_\mathrm{\bot
R}(\phi)~ R\left(1-\beta \cos(\phi/2) \right) d\phi =
\nonumber\\&&\cr = { e^2 \lambda_0\over {4 \pi
\varepsilon_\mathrm{0} R}} \left[\beta^2
\ln(\tan(\phi/2))+{\beta(1-\beta^2)\over{ 1-\beta\cos(\phi/2)}}
\right]_\mathrm{\phi_\mathrm{min}}^{\phi_\mathrm{max}}&& ~,
\label{intR}
\end{eqnarray}
where the superscript "B" stands for "bunch", and $\lambda_0$ is
the constant linear density of the bunch. If we now expand in
$\phi_\mathrm{min} \ll 1$ and $\phi_\mathrm{max} \ll 1$ the
trigonometric functions in Eq. (\ref{intC}) and Eq. (\ref{intR})
we have
\begin{eqnarray}
{F_\mathrm{\bot C}^\mathrm{B}} \simeq  { e^2 \lambda_0 \over {4
\pi \varepsilon_\mathrm{0} R}} \Bigg[-{1\over{\gamma^2}}
\ln\left({\phi_\mathrm{max}\over{\phi_\mathrm{min}}}\right)\nonumber\\&&\cr
-{1\over{2 \gamma^2}}\left(
{1\over{1-\beta+\beta\phi_\mathrm{max}^2/8}} -
{1\over{1-\beta+\beta\phi_\mathrm{min}^2/8}}\right)
\Bigg]&&~\label{expBC}
\end{eqnarray}
and
\begin{eqnarray}
{F_\mathrm{\bot R}^\mathrm{B}} \simeq  { e^2 \lambda_0 \over {4
\pi \varepsilon_\mathrm{0} R}} \Bigg[
\ln\left({\phi_\mathrm{max}\over{\phi_\mathrm{min}}}\right)\nonumber\\&&\cr
+{1\over{\gamma^2}}\left(
{1\over{1-\beta+\beta\phi_\mathrm{max}^2/8}} -
{1\over{1-\beta+\beta\phi_\mathrm{min}^2/8}}\right)
\Bigg]&&~.\label{expBR}
\end{eqnarray}
The second term on the right side of Eq. (\ref{expBC}) is
centrifugal, as well as the first one on the right side of Eq.
(\ref{expBR}). The other ones are centripetal. Note that the
logarithmic term in Eq. (\ref{expBC}) is unimportant (in the limit
of large values for $\gamma$) with respect to the one in Eq.
(\ref{expBR}), while the second term in Eq. (\ref{expBC}) modifies
for a factor $1/2$ the analogous centripetal term in Eq.
(\ref{expBR}). Therefore, the total transverse force on the test
electron is given by:
\begin{equation}
{F_\mathrm{\bot}^\mathrm{B}} \simeq  { e^2 \lambda_0 \over {4 \pi
\varepsilon_\mathrm{0} R}} \left[
\ln\left({\hat{\phi}_\mathrm{max}\over{\hat{\phi}_\mathrm{min}}}\right)
+\left( {4\over{4+\hat{\phi}_\mathrm{max}^2}} -
{4\over{4+\hat{\phi}_\mathrm{min}^2}}\right)\right]~,\label{expBtot}
\end{equation}
which is the sum of a logarithmic centrifugal term and a
centripetal term. It is easy to check that Eq. (\ref{expBtot})
can be also obtained by direct integration of Eq.(\ref{Totexp}).

A natural assumption is to consider our bunch density such that
there are many particles within a distance $R/\gamma^3$ which, as
it has already been underlined, is the minimal characteristic
distance of the system: if the test particle is to be considered
at the head of our bunch, then it is straightforward to assume
$\Delta s_\mathrm{min} \ll R/\gamma^3$ which, by means of the
retardation condition Eq. (\ref{retsteadyappr}), gives us back
the condition: $\phi_\mathrm{min} \ll 1/\gamma$, the non-linear
term in $\phi_\mathrm{min}$ of Eq. (\ref{retsteadyappr}) being,
in this case, negligible.

Under the latter assumption we can easily investigate the two
cases of a short bunch, that is $\phi_\mathrm{max} \ll 1/\gamma$
(in which the linear term in $\phi_\mathrm{max}$ of Eq.
(\ref{retsteadyappr}) dominates), and of a long bunch, that is
$\phi_\mathrm{max} \gg 1/\gamma$ (in which the linear term in
$\phi_\mathrm{max}$ of Eq. (\ref{retsteadyappr}) is negligible).
These two cases correspond, of course, to the asymptotic
situations for the two-particle interaction discussed in Section
\ref{sec:steadytwo}.

Let us consider first the case $\phi_\mathrm{max} \ll 1/\gamma$.
Eq. (\ref{expBC}), Eq. (\ref{expBR}) and Eq. (\ref{expBtot}),
then, read

\begin{eqnarray}
{F_\mathrm{\bot C}^\mathrm{B}} \simeq  { e^2 \lambda_0 \over {4
\pi \varepsilon_\mathrm{0} R}} \left[ -{1\over{\gamma^2}}
\ln\left({\hat{\phi}_\mathrm{max}\over{\hat{\phi}_\mathrm{min}}}\right)
+ {1\over{4}}\hat{\phi}^2_\mathrm{max}\right]\nonumber\\&&\cr
\simeq { e^2 \lambda_0 \over {4 \pi \varepsilon_\mathrm{0} R}}
\left[ -{1\over{\gamma^2}}\ln\left({\Delta
\hat{s}_\mathrm{max}\over{\Delta \hat{s}_\mathrm{min}}}\right)+
\Delta \hat{s}_\mathrm{max}^2 \right]~, &&\label{shortBC}
\end{eqnarray}
\begin{eqnarray}
{F_\mathrm{\bot R}^\mathrm{B}} \simeq  { e^2 \lambda_0 \over {4
\pi \varepsilon_\mathrm{0} R}} \left[
\ln\left({\hat{\phi}_\mathrm{max}\over{\hat{\phi}_\mathrm{min}}}\right)
- {1\over{2}}\hat{\phi}^2_\mathrm{max}\right] \nonumber\\&&\cr
\simeq { e^2 \lambda_0 \over {4 \pi \varepsilon_\mathrm{0} R}}
\left[\ln\left({\Delta \hat{s}_\mathrm{max}\over{\Delta
\hat{s}_\mathrm{min}}}\right)-2 \Delta \hat{s}_\mathrm{max}^2
\right] && \label{shortBR}
\end{eqnarray}
and
\begin{eqnarray}
{F_\mathrm{\bot}^\mathrm{B}} \simeq  { e^2 \lambda_0 \over {4 \pi
\varepsilon_\mathrm{0} R}} \left[
\ln\left({\hat{\phi}_\mathrm{max}\over{\hat{\phi}_\mathrm{min}}}\right)
- {1\over{4}}\hat{\phi}^2_\mathrm{max}\right] \nonumber\\&&\cr
\simeq { e^2 \lambda_0 \over {4 \pi \varepsilon_\mathrm{0} R}}
\left[\ln\left({\Delta \hat{s}_\mathrm{max}\over{\Delta
\hat{s}_\mathrm{min}}}\right)- \Delta \hat{s}_\mathrm{max}^2
\right]~. && \label{shortBtot}
\end{eqnarray}
From Eq. (\ref{shortBtot}) we see that the centripetal term tends,
asymptotically, to zero as $(\gamma \phi_\mathrm{max})^2$.

In the case $\phi_\mathrm{max} \gg 1/\gamma$, instead, we have

\begin{eqnarray}
{F_\mathrm{\bot C}^\mathrm{B}} \simeq  { e^2 \lambda_0 \over {4
\pi \varepsilon_\mathrm{0} R}} \left[-{1\over{\gamma^2}}
\ln\left({\hat{\phi}_\mathrm{max}\over{\hat{\phi}_\mathrm{min}}}\right)
+ 1 \right] \nonumber\\&&\cr \simeq { e^2 \lambda_0 \over {4 \pi
\varepsilon_\mathrm{0} R}} \left[ -{1\over{\gamma^2}}\ln\left({(
24 \Delta \hat{s}_\mathrm{max})^{1/3}\over{\Delta
\hat{s}_\mathrm{min}}}\right)+1 \right]&&~, \label{longBC}
\end{eqnarray}
\begin{eqnarray}
{F_\mathrm{\bot R}^\mathrm{B}} \simeq  { e^2 \lambda_0 \over {4
\pi \varepsilon_\mathrm{0} R}} \left[
\ln\left({\hat{\phi}_\mathrm{max}\over{\hat{\phi}_\mathrm{min}}}\right)
- 2 \right] \nonumber\\&&\cr \simeq { e^2 \lambda_0 \over {4 \pi
\varepsilon_\mathrm{0} R}} \left[ \ln\left({( 24 \Delta
\hat{s}_\mathrm{max})^{1/3}\over{\Delta
\hat{s}_\mathrm{min}}}\right)-2 \right]&& \label{longBR}
\end{eqnarray}
and
\begin{eqnarray}
{F_\mathrm{\bot}^\mathrm{B}} \simeq  { e^2 \lambda_0 \over {4 \pi
\varepsilon_\mathrm{0} R}} \left[
\ln\left({\hat{\phi}_\mathrm{max}\over{\hat{\phi}_\mathrm{min}}}\right)
- 1 \right] \nonumber\\&&\cr \simeq { e^2 \lambda_0 \over {4 \pi
\varepsilon_\mathrm{0} R}} \left[ \ln\left({( 24 \Delta
\hat{s}_\mathrm{max})^{1/3}\over{\Delta
\hat{s}_\mathrm{min}}}\right)-1 \right]~,&& \label{longBtot}
\end{eqnarray}
which means that the centripetal term saturates to a constant
value in the limit of a long bunch. The case of a coasting beam
(which fits our particular case $\hat{\phi}_\mathrm{max} \gg 1$)
with transverse extent $h \gg R/\gamma^2$ has been discussed in
literature (see \cite{RUI2}, \cite{DERBT}, \cite{STUP}). As
already mentioned above, the presence, in the choice of the model
made in the latter works, of a non-negligible transverse dimension
alters the structure of the retardation condition (with respect
to the one given in Eq. (\ref{retsteadyappr})), which explains
why the centrifugal part of the transverse force scales, in
literature, as $\ln(R/h)$. Moreover, again because of the
presence of a transverse dimension, the velocity contribution to
the transverse force can be discarded giving the same centripetal
term which we found from the radiative term only in Eq.
(\ref{longBR}), that is $ -{2 e^2 \lambda_0 / (4 \pi
\varepsilon_\mathrm{0} R)}$.

Note that, in order to obtain Eq. (\ref{longBR}), we had to
integrate contributions from $\hat{\phi} \ll 1$ to $\hat{\phi} \gg
1$: we can give a simple explanation for the centripetal constant
force term just analyzing Eq. (\ref{Totexp}) and  its asymptotic
behaviours. The product of Eq. (\ref{Totexp}) with the Jacobian
(Eq. (\ref{Jacobian})) of the transformation between $\Delta
\hat{s}$ and $\hat{\phi}$ (this product is just the integrand in
Eq. (\ref{intR})) is equal, in the limits for $\hat{\phi} \gg 1$
and $\hat{\phi} \ll 1$, to $1/\hat{\phi}$: then we can conclude
that the logarithmic centrifugal term in Eq. (\ref{longBR}) (as
well as in Eq. (\ref{longBC} and Eq. (\ref{longBtot}))  takes
into account the $1/\hat{\phi}$-behaviour of the transverse force
for the two-particle system, while the constant centripetal term
brings information about the way in which the transverse force
for a two-particle system changes in the intermediate region
between the limits $\hat{\phi} \ll 1$ and $\hat{\phi} \gg 1$.
Note, however, that there is no physical ground to distinguish
between the first (centrifugal) and the second (centripetal) term
in Eq. (\ref{longBtot}): the total force is always centrifugal,
and both terms are consequences of the integration of a unique
expression for the force between a two-particle system (which, of
course, is always centrifugal too).

\section{\label{sec:transientwo}TRANSVERSE INTERACTION BETWEEN
TWO ELECTRONS MOVING IN AN ARC OF A CIRCLE}

\begin{figure}[htb]
\includegraphics*[width=90mm]{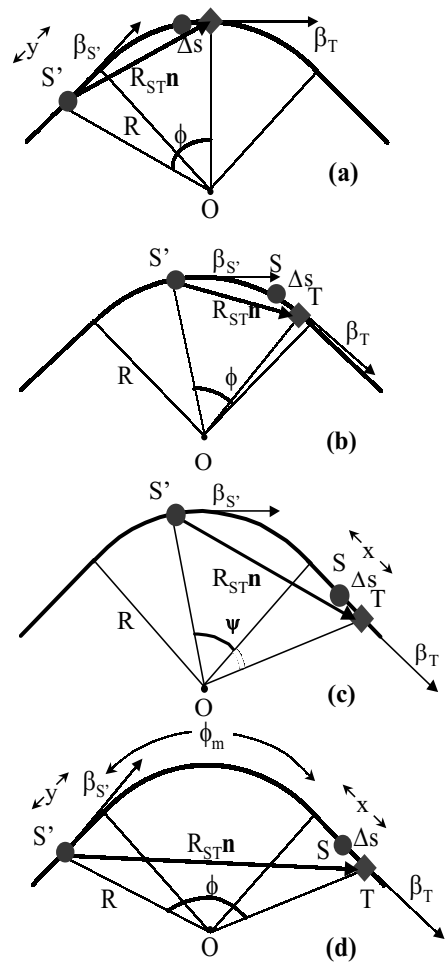}
\caption{\label{FIG5} Relative configuration of the retarded
source point S'and the test point T for a system of two electron
passing a bending magnet.}
\end{figure}

We will now discuss the case of a two-particle system during the
passage from a straight path to a circular one and from a
circular path to a straight one. The four possible cases are
sketched in Fig. \ref{FIG5} for the case in which the test
particle is in front of the source.

The case in which both particles are in the bend, depicted in Fig.
\ref{FIG5}b, has already been discussed in Section
\ref{sec:steadytwo}. Moreover, note that the situation in which
the source particle is ahead of the test electron can be treated
immediately for all three (a, c and d) transient cases in Fig.
\ref{FIG5} (of course, with respect to the figure, test and source
particle exchange roles) on the basis of Eq. (\ref{Fperpht}). In
fact in such a case, we can assume the retarded angle $\phi$
small enough (the test particle "runs against" the
electromagnetic signal) so that the actual trajectory followed by
the particles is not essential and one can use Eq.
(\ref{Fperpht}) to describe also the transient cases. Now, the
important contribution from the source particle comes from the
acceleration part of the Lienard-Wiechert fields. Then, within
our approximations, the only non-negligible contribution is
constant and identical to the one in Eq. (\ref{Fperpht}), and it
is present in the situation (again, with the roles of test and
source particle inverted) depicted in Fig. \ref{FIG5}a only (Fig.
\ref{FIG5}b is just the steady-state case).

We will discuss more extensively the consequences of this fact in
Section \ref{sec:transientbunch}.

Let us now focus on the cases in which the source particle is
behind the test particle and, in particular, let us  first deal
with the case in Fig. \ref{FIG5}a; such a situation occurs when
the following condition is met (see \cite{SALLONG}):

\begin{equation}
\Delta \hat{s}> {\hat{\phi}\over{2}}+{\hat{\phi}^3\over{24}}.
\label{cond:a}
\end{equation}
Let us designate with $y$ the distance, along the straight line
before the bend, between the retarded source particle and the
beginning of the magnet. The retardation condition, in its
approximated form, reads (see \cite{SALLONG})

\begin{equation}
\Delta \hat{s} \simeq
{\hat{\phi}+\hat{y}\over{2}}+{\hat{\phi}^3\over{24}}{\hat{\phi}+
4\hat{y}\over{\hat{\phi}+\hat{y}}}, \label{retcond:a}
\end{equation}
where we introduced the normalized quantity $\hat{y} = y
\gamma/R$, which is  just $y/R$ normalized to the synchrotron
radiation formation angle, $1/\gamma$.

In the situation considered, the source particle is only
responsible for a velocity contribution, therefore
$F_\mathrm{\bot} = F_\mathrm{\bot C}$. By direct use of Eq.
(\ref{Coulomb}), one can find the exact expression for
$F_\mathrm{\bot}$

\begin{widetext}
\begin{equation}
F_\mathrm{\bot} = { e^2  \over {4 \pi \varepsilon_\mathrm{0}
\gamma^2}}{\beta^2 R (1-\cos\phi) - \beta \sin\phi [(y+R
\sin\phi)^2 + R^2 (1-\cos\phi) ]^{1/2} + R - R\cos\phi+y \sin\phi
\over{\left\{[(y+R \sin\phi)^2 + R^2 (1-\cos\phi) ]^{1/2} - \beta
y - R \beta \sin\phi )^3\right\}}} \label{triga}
\end{equation}
\end{widetext}
\begin{figure*}
\includegraphics*[width=180mm]{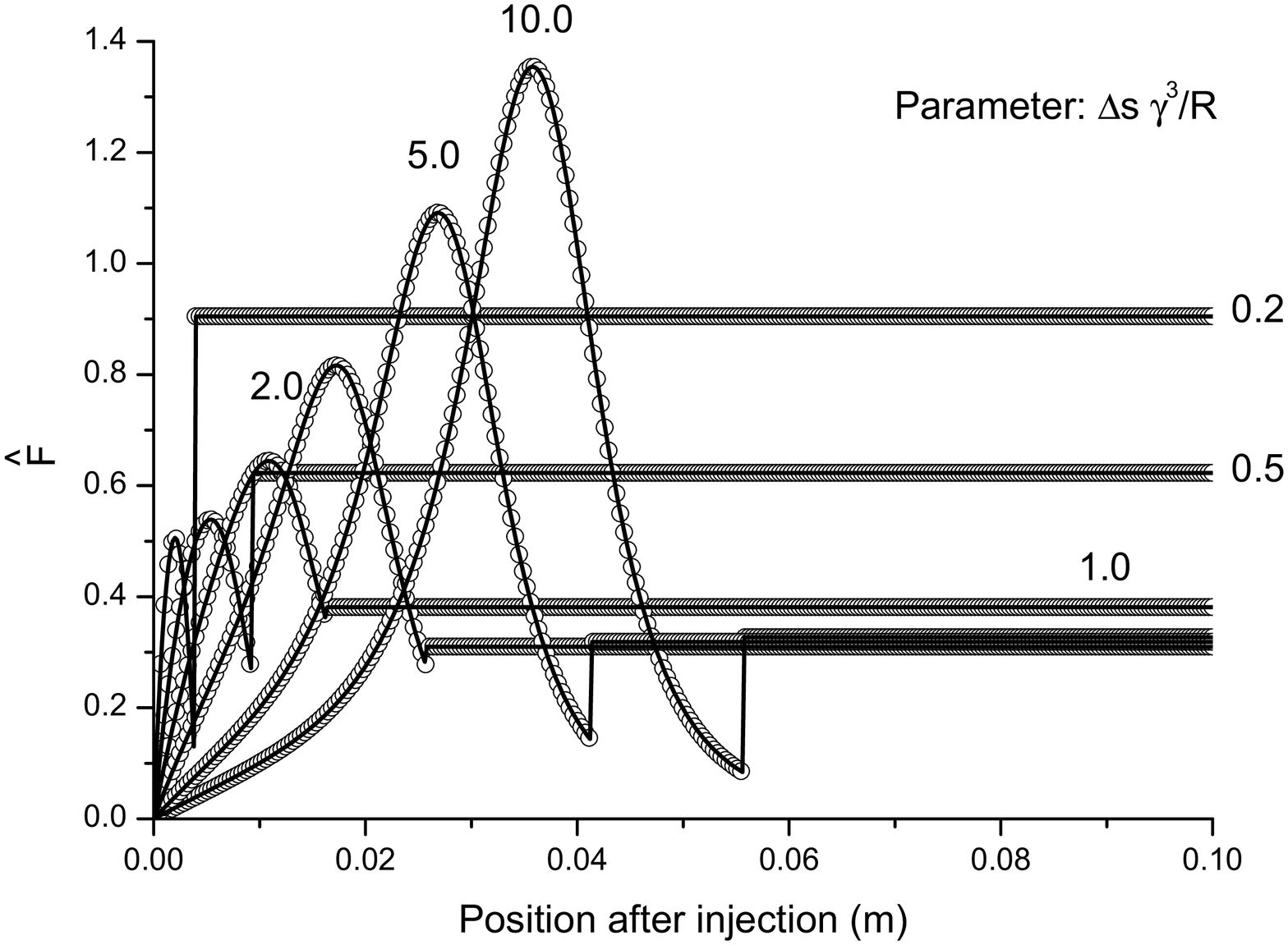}
\caption{\label{FIG6} Normalized transverse force ($\hat{F} =
F_\mathrm{\bot}/[e^2/(4\pi\varepsilon_0\Delta s)]$) for a
two-particle system entering a hard-edge bending magnet as a
function of the position after injection. The solid lines show
analytical results; the circles describe the outcome from
$\mathrm{TRAFIC}^4$. We plotted several outcomes from different
values of the normalized distance between the two particles. }
\end{figure*}

Expanding the trigonometric functions in Eq. (\ref{triga}) and
using normalized quantities one finds:

\begin{eqnarray}
F_\mathrm{\bot} \simeq { e^2  \over {4 \pi \varepsilon_\mathrm{0}
}} {4 \gamma^3
 \over {R^2}} \hat{\phi}(\hat{y}+\hat{\phi})^2 \nonumber\\&&\cr \times {\hat{y}^2
+\hat{y}\left( \hat{\phi}+\hat{\phi}^3/2\right)+{\hat{\phi}^4/4}
\over{\left[(\hat{y}+\hat{\phi})^2+{\hat{\phi}^4/4}\right]^3}}&&
\label{expa}
\end{eqnarray}
It can be easily verified that, as it must be, Eq. (\ref{expa})
reduces to Eq. (\ref{Coulexp}) in the limit $y \rightarrow 0$.

It is now possible, by means of Eq. (\ref{expa}), to plot the
normalized transverse force $\hat{F} = F_\mathrm{\bot}/
 [e^2/(4\pi\varepsilon_0\Delta s)]$ as a function of
the position after the injection (defined by the entrance of the
test particle in the hard-edge magnet) for different values of
$\Delta \hat{s} = \Delta s \gamma^3/R$. In Fig. \ref{FIG6} we
compared such a plot with numerical results from the code
$\mathrm{TRAFIC}^4$ (see \cite{ROTH}).

Note that, at the position which corresponds to the entrance of
the retarded source in the magnet there is a discontinuity in the
plots. This is linked to our model choice, and it is due to the
abrupt (hard edge magnet) switching on of the acceleration fields.

As general remark to Fig. \ref{FIG6} (and to the following ones)
it might be worthwhile to stress that the perfect agreement (with
graphical accuracy) between our calculations and numerical
results by $\mathrm{TRAFIC}^4$ provides, \textit{per se}, an
excellent cross-check between analysis and simulations, which
enhance one's level of confidence on both these approaches.

Let us now consider the case depicted in Fig. \ref{FIG5}c, in
which the source particle has its retarded position inside the
bend and the test particle has its present position in the
straight line following the magnet. We will define with $x$ the
distance, along the straight line after the magnet, between the
end of the bend and the present position of the test particle. In
this situation the following condition is verified (see
\cite{SALLONG}):

\begin{equation}
\Delta \hat{s} <
{\hat{\phi}_\mathrm{m}+\hat{x}\over{2}}+{\hat{\phi}_\mathrm{m}^3\over{24}}{\hat
{\phi}_\mathrm{m}+4\hat{x}\over{\hat{\phi}_\mathrm{m}+\hat{x}}},
\label{cond:b}
\end{equation}
where $\hat{\phi}_\mathrm{m} = \gamma \phi_\mathrm{m}$,
$\phi_\mathrm{m}$ being the angular extension of magnet, and
$\hat{x}=\gamma x/R$ (the reason for this normalization choice
for $x$ is identical to that for $y$) .

The retardation condition reads

\begin{equation}
\Delta \hat{s} \simeq
{\hat{\psi}+\hat{x}\over{2}}+{\hat{\psi}^3\over{24}}{\hat{\psi}+
4\hat{x}\over{\hat{\psi}+\hat{x}}}. \label{retcond:b}
\end{equation}
In contrast with the case of Fig. \ref{FIG5}a, here we have
contributions from both velocity and acceleration field. Again, by
direct use of Eq. (\ref{Coulomb}) and Eq. (\ref{Radiation}) one
can find the exact expression for the transverse electromagnetic
force performed by the source particle on the test particle

\begin{equation}
F_\mathrm{\bot} = F_\mathrm{\bot C} + F_\mathrm{\bot R}~,
\label{Fperptotb}
\end{equation}
where
\begin{widetext}
\begin{equation}
F_\mathrm{\bot C} = { e^2  \over {4 \pi \varepsilon_\mathrm{0}
\gamma^2}} {R (1-\cos\psi)(1-\beta^2 \cos\psi)-\beta\sin\psi
\left[\left((x + R\sin\psi)^2+R^2(1-\cos\psi)^2
\right)^{1/2}-\beta x -\beta R\sin\psi
\right]\over{\left[\left((x + R\sin\psi)^2+R^2(1-\cos\psi)^2
\right)^{1/2}-\beta x \cos\psi -\beta R \sin\psi \right]^3}}~
\label{FperpbC}
\end{equation}
and
\begin{eqnarray}
F_\mathrm{\bot R} = { e^2  \over {4 \pi \varepsilon_\mathrm{0} }}
{\beta^2 \over{R}}\Bigg\{ {- \psi\left[ \left((x +
R\sin\psi)^2+R^2(1-\cos\psi)^2 \right)^{1/2} + \beta x +\beta R
\sin \psi \right] - \beta R (1- \cos\psi) \sin\psi
\over{\left[\left((x + R\sin\psi)^2+R^2(1-\cos\psi)^2
\right)^{1/2}-\beta x \cos\psi -\beta R \sin\psi \right]^2
}}+\nonumber\\&&\cr + {\left( R+x \sin\psi - R \cos\psi\right)
\over{\left[\left((x + R\sin\psi)^2+R^2(1-\cos\psi)^2
\right)^{1/2}-\beta x \cos\psi -\beta R \sin\psi \right]^3}}
\times \nonumber\\&&\cr \times \Bigg[
R(1-\cos\psi)(1-\beta^2\cos\psi)-\beta\sin\psi\left[\left((x +
R\sin\psi)^2+R^2(1-\cos\psi)^2 \right)^{1/2} - \beta x - \beta R
\sin\psi \right] \Bigg] \Bigg\} && .\label{FperpbR}
\end{eqnarray}
\end{widetext}
Expanding the trigonometric functions in Eq. (\ref{FperpbC}) and
Eq. (\ref{FperpbR}), and using normalized quantities one finds:

\begin{figure*}
\includegraphics*[width=180mm]{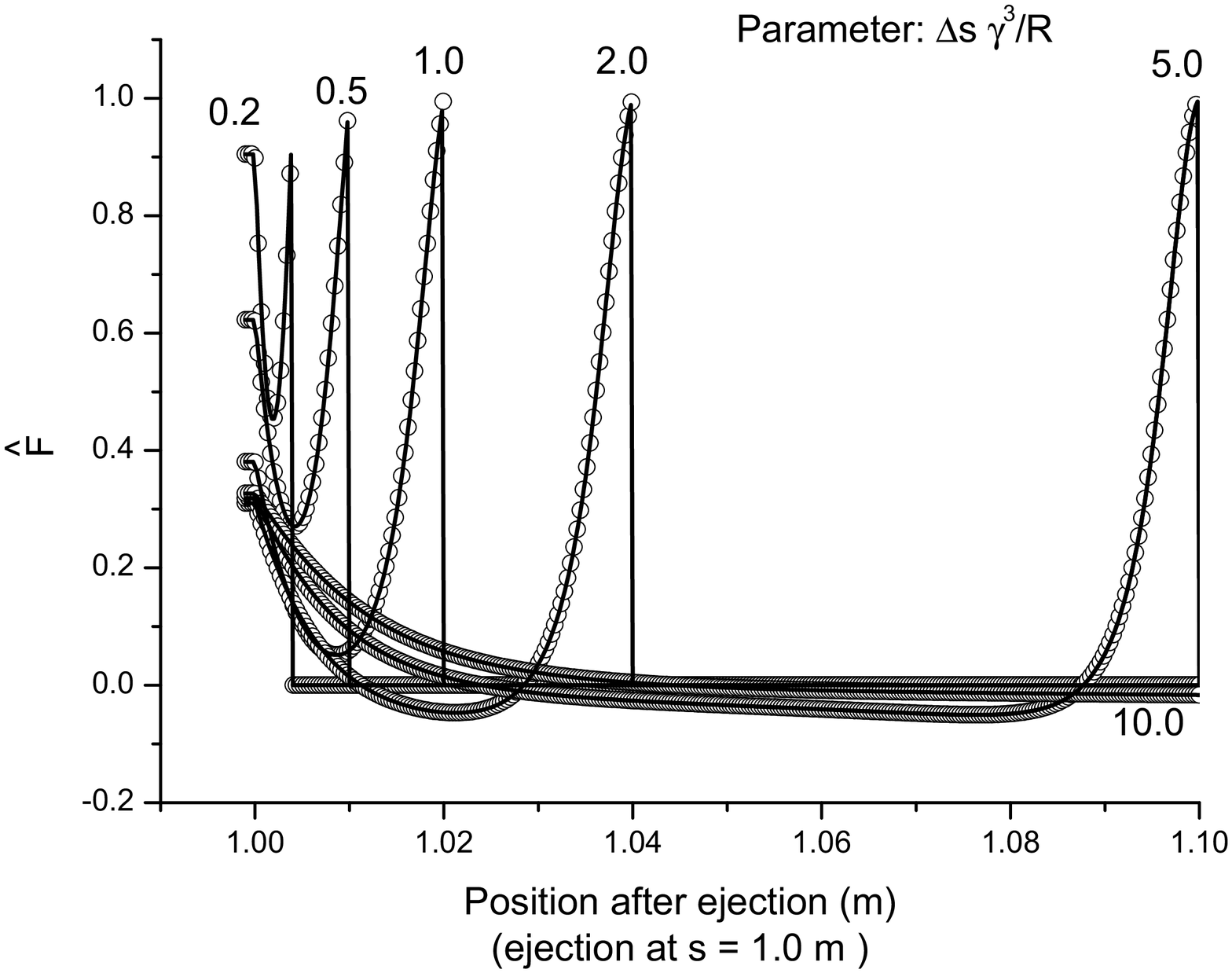}
\caption{\label{FIG7} Normalized transverse force ($\hat{F} =
F_\mathrm{\bot}/[e^2/(4\pi\varepsilon_0\Delta s)]$) for a
two-particle system leaving a hard-edge bending magnet as a
function of the position after the ejection. The solid lines show
analytical results; the circles describe the outcome from
$\mathrm{TRAFIC}^4$. We plotted several outcomes from different
values of the normalized distance between the two particles. }
\end{figure*}
\begin{widetext}
\begin{equation}
F_\mathrm{\bot C} = {2 e^2 \gamma^3 \over {4 \pi
\varepsilon_\mathrm{0} R^2}} \hat{\psi}(\hat{x}+\hat{\psi})^2
{-2\hat{x}^2
+\hat{x}\left(\hat{\psi}^3
-2\hat{\psi}\right)+{\hat{\psi}^4/2}
\over{\left[(\hat{x}+ \hat{\psi})^2 +
({\hat{\psi}^2/4})(2\hat{x}+\hat{\psi})^2 \right]^3 }} ~,
\label{apprFbC}
\end{equation}
and

\begin{equation}
F_\mathrm{\bot R} = {2 e^2 \gamma^3 \over {4 \pi
\varepsilon_\mathrm{0} R^2}} (\hat{x}+ \hat{\psi}) \left\{
{\hat{x}^2+\hat{x}\hat{\psi}(2-\hat{\psi}^2)+\hat{\psi}^2- 3/4
\hat{\psi}^4 \over{\left[(\hat{x} + \hat{\psi})^2 +
(\hat{\psi}^2/4)(2\hat{x}+\hat{\psi})^2 \right]^2 }} +
{(\hat{x}+\hat{\psi})(\hat{x}+{\hat{\psi}/2})\hat{\psi}^2\left[
-2 \hat{x}^2+\hat{x}\hat{\psi}(-2+\hat{\psi}^2)+
{\hat{\psi}^4/2}\right] \over{\left[(\hat{x} + \hat{\psi})^2 +
(\hat{\psi}^2/4)(2\hat{x}+\hat{\psi})^2 \right]^3}} \right\}~.
\label{apprFbR}
\end{equation}
\end{widetext}
Similarly to the latter case, it can be easily verified that Eq.
(\ref{apprFbC}) and Eq. (\ref{apprFbR}) reduce to Eq.
(\ref{Coulexp}) and Eq. (\ref{Radexp}), respectively,  in the
limit $x \rightarrow 0$. Again, it is possible to plot the
normalized transverse force $\hat{F}$ (defined above) as a
function of the position after the ejection (defined by the exit
of the test particle from the hard-edge magnet) for different
values of $\Delta \hat{s} = \Delta s \gamma^3/R$. In Fig.
\ref{FIG7} we compared such a plot with numerical results from
$\mathrm{TRAFIC}^4$.

Again, at the position which corresponds to the exit of the
retarded source from the magnet there is a discontinuity in the
plots. This, again, is linked to our model choice, and it is due
to the fact that, for particles on axis, when the retarded source
leaves the magnet there is only Coulomb repulsion along the
longitudinal direction.

The last case left to discuss is depicted in Fig. \ref{FIG5}d;
the source particle has its retarded position in the straight line
before the bend, and the test particle has its present position
in the straight line following the magnet. This case occurs when

\begin{equation}
\Delta \hat{s} >
{\hat{\phi}_\mathrm{m}+\hat{x}\over{2}}+{\hat{\phi}_\mathrm{m}^3\over{24}}{\hat
{\phi}_\mathrm{m}+4\hat{x}\over{\hat{\phi}_\mathrm{m}+\hat{x}}}.
\label{cond:c}
\end{equation}

The retardation condition reads

\begin{equation}
\Delta \hat{s} \simeq
{\hat{\phi}_\mathrm{m}+\hat{x}+\hat{y}\over{2}}+{\hat{\phi}_\mathrm{m}^3\over{24}}
{\hat{\phi}_\mathrm{m}+
4\hat{\phi}_\mathrm{m}(\hat{x}+\hat{y})\over{\hat{\phi}_\mathrm{m}+\hat{x}+\hat{y}}}.
\label{retcond:c}
\end{equation}

In this case we have only velocity contributions. The exact
expression for the electromagnetic transverse force on the test
particle is

\begin{figure*}
\includegraphics*[width=180mm]{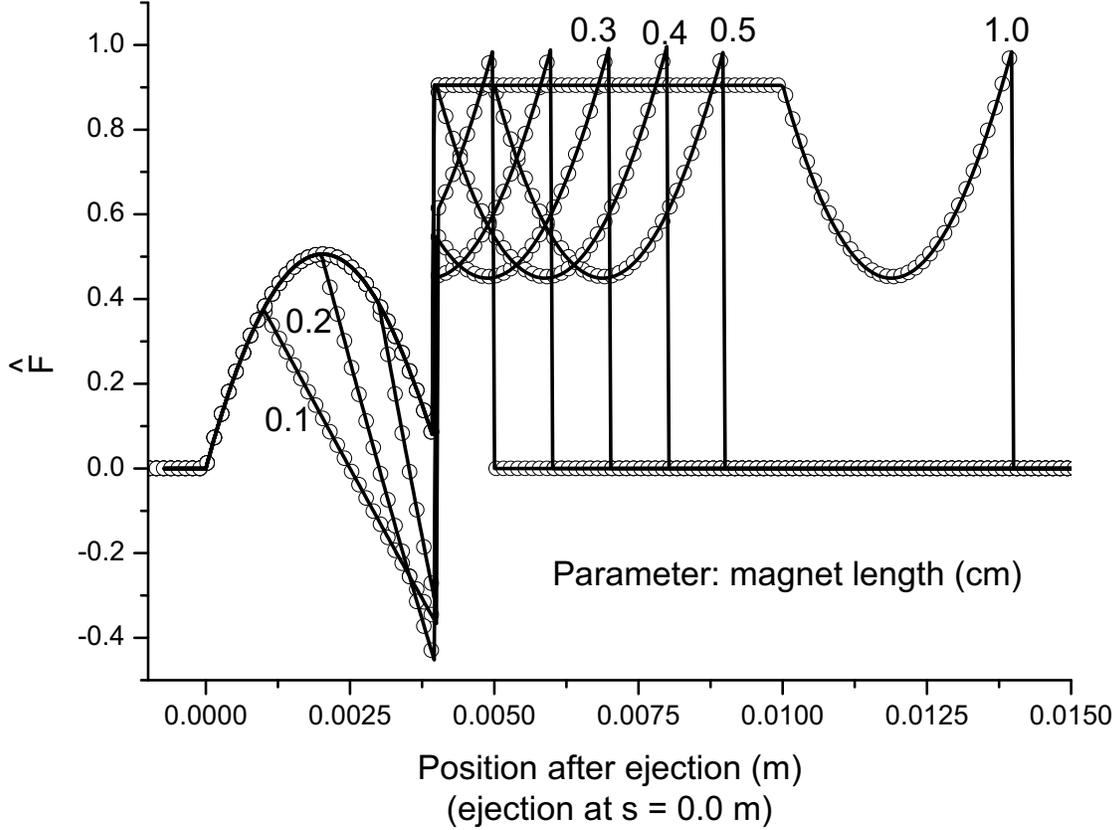}
\caption{\label{FIG8} Normalized transverse force ($\hat{F} =
F_\mathrm{\bot}/[e^2/(4\pi\varepsilon_0\Delta s)]$) for a
two-particle system crossing a hard-edge bending magnet as a
function of the position of the test particle inside the magnet
in the case of a short magnet $\hat{\phi}_\mathrm{m} \ll 1$. The
solid lines show analytical results; the circles describe the
outcome from $\mathrm{TRAFIC}^4$. Here the normalized distance
between the two particles is $\Delta \hat{s} = 0.2$. We plotted
several outcomes from different values of the magnet length $R
\hat{\phi}_\mathrm{m} = 0.1,~ 0.2,~ 0.3,~ 0.4,~ 0.5,~ 1.0$. }
\end{figure*}
\begin{figure*}
\includegraphics*[width=180mm]{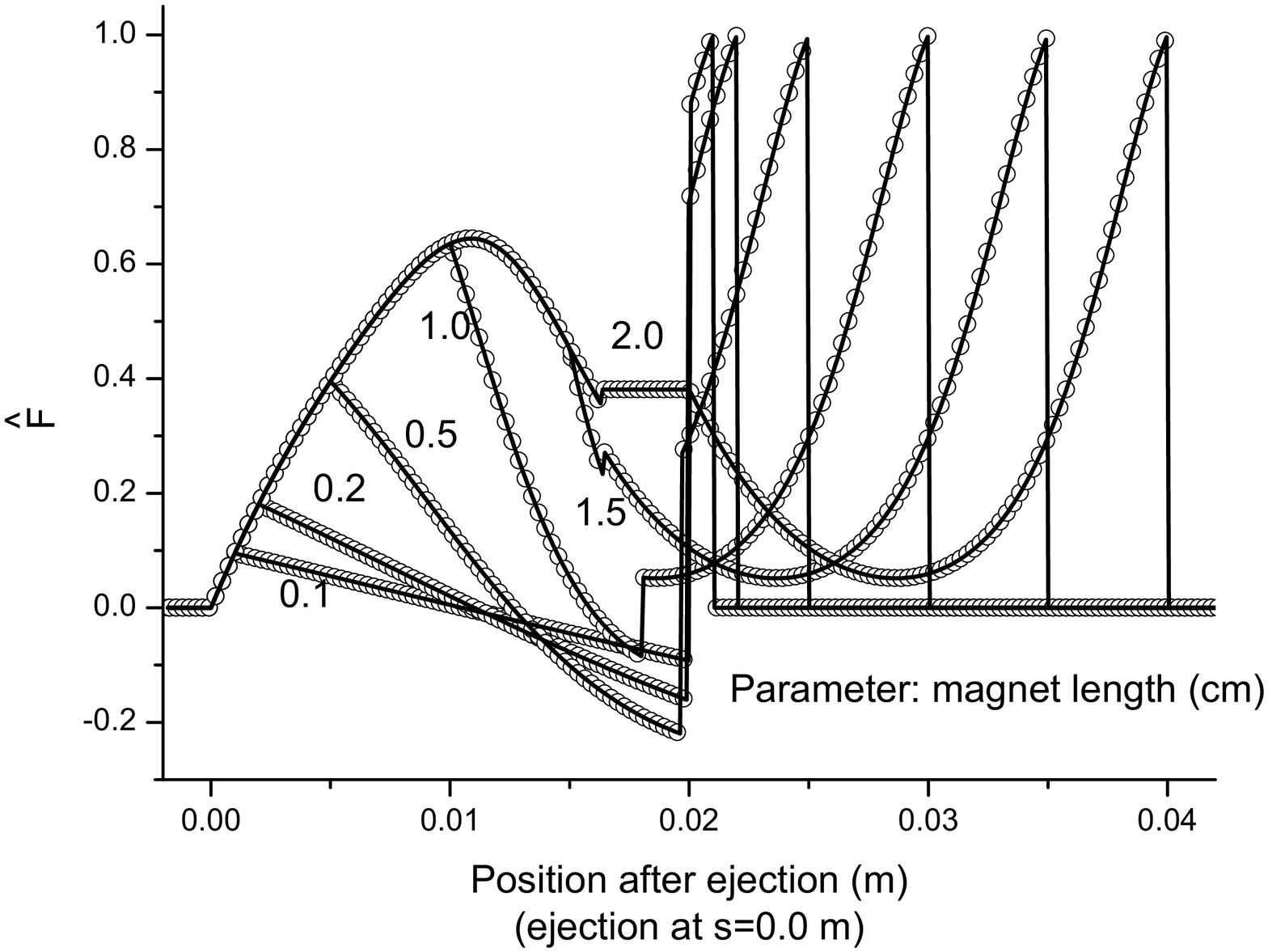}
\caption{\label{FIG9} Normalized transverse force ($\hat{F} =
F_\mathrm{\bot}/[e^2/(4\pi\varepsilon_0\Delta s)]$) for a
two-particle system crossing a hard-edge bending magnet as a
function of the position of the test particle inside the magnet
in the case of a short magnet $\hat{\phi}_\mathrm{m} \ll 1$. The
solid lines show analytical results; the circles describe the
outcome from $\mathrm{TRAFIC}^4$. Here the normalized distance
between the two particles is $\Delta \hat{s} = 1.0$. We plotted
several outcomes from different values of the magnet length $R
\hat{\phi}_\mathrm{m} = 0.1,~ 0.2,~ 0.5,~ 1.0,~ 1.5,~ 2.0$. }
\end{figure*}
\begin{figure*}
\includegraphics*[width=180mm]{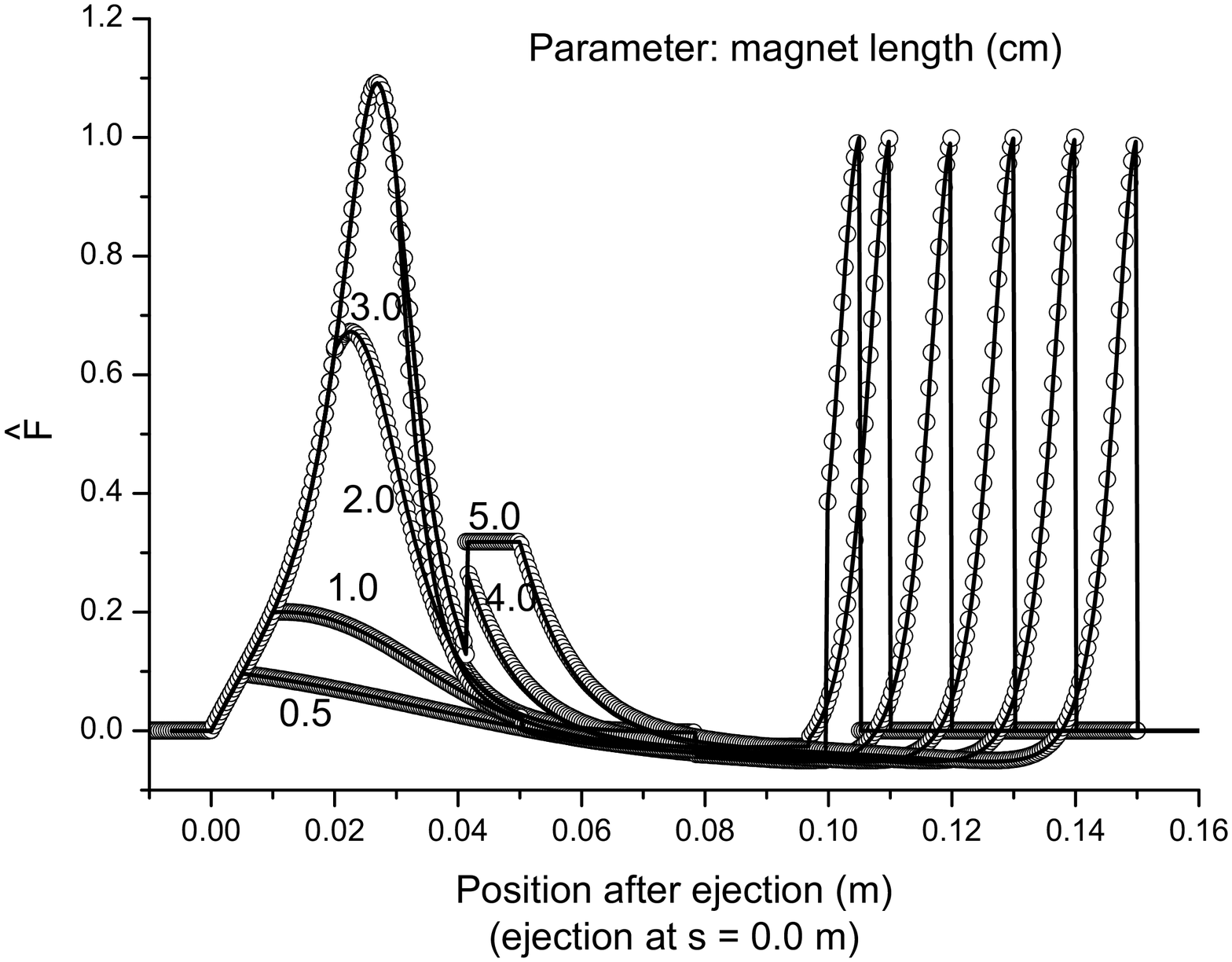}
\caption{\label{FIG10} Normalized transverse force ($\hat{F} =
F_\mathrm{\bot}/[e^2/(4\pi\varepsilon_0\Delta s)]$) for a
two-particle system crossing a hard-edge bending magnet as a
function of the position of the test particle inside the magnet
in the case of a short magnet $\hat{\phi}_\mathrm{m} \ll 1$.The
solid lines show analytical results; the circles describe the
outcome from $\mathrm{TRAFIC}^4$. Here the normalized distance
between the two particles is $\Delta \hat{s} = 5.0$. We plotted
several outcomes from different values of the magnet length $R
\hat{\phi}_\mathrm{m} = 0.5,~ 1.0,~ 2.0,~ 3.0,~ 4.0,~ 5.0$. }
\end{figure*}

\begin{widetext}
\begin{eqnarray}
F_\mathrm{\bot} = {e^2 \over {4 \pi \varepsilon_\mathrm{0}
\gamma^2}} \Big\{
(R+y\sin\phi_\mathrm{m}-R\cos\phi_\mathrm{m})(1-\beta^2\cos\phi_\mathrm{m})-\beta\sin\phi_\mathrm{m}\Big[\big(2
R^2 + x^2 + y^2 + \cos\phi_\mathrm{m}(2 x y - 2 R^2)
\nonumber\\&&\cr+ 2 R (x+y)
\sin\phi_\mathrm{m}\big)^{1/2}-\beta(x+y\cos\phi_\mathrm{m}+R\sin\phi_\mathrm{m})\Big]\Big\}\Big\{\left[2
R^2 + x^2 + y^2 + 2\cos\phi_\mathrm{m}(x y - R^2) + 2 R (x+y)
\sin\phi_\mathrm{m}\right]^{1/2}\nonumber\\&&\cr
-\beta\cos\phi_\mathrm{m}(x+y\cos\phi_\mathrm{m}+R\sin\phi_\mathrm{m})-\beta\sin\phi_\mathrm{m}(R+y\sin\phi_\mathrm{m}
-R\cos\phi_\mathrm{m})\Big\}^{-3} ~. && \label{Fperpc}
\end{eqnarray}
\end{widetext}
Expanding the trigonometric functions in Eq. (\ref{Fperpc}) and
using normalized quantities one finds the following approximated
expression for $F_\mathrm{\bot}$:
\begin{widetext}
\begin{eqnarray}
F_\mathrm{\bot} \simeq {e^2 \over {4 \pi \varepsilon_\mathrm{0}
R^2}} 8 \gamma^3
(\hat{x}+\hat{y}+\hat{\phi_\mathrm{m}})^2\phi_\mathrm{m} \left\{
-{\hat{x}^2/2}+
{\hat{y}^2/2}
+ ({\hat{\phi}_\mathrm{m}^2/2})\hat{x}\hat{y}
+{\hat{x}(\hat{\phi}_\mathrm{m}^3/4 -\hat{\phi}_\mathrm{m}/2)}
+{\hat{y}(\hat{\phi}_\mathrm{m}^3/4+\hat{\phi}_\mathrm{m}/2)} +
{\hat{\phi}_\mathrm{m}^4/8}\right\}\nonumber\\&&\cr \times{\left\{
(\hat{x}+\hat{y}+\hat{\phi}_\mathrm{m})\left[\hat{x}
(1+\hat{\phi}^2_\mathrm{m}) +
\hat{y}+\hat{\phi}_\mathrm{m}+{\hat{\phi}^3_\mathrm{m}
/3}\right]-(\hat{\phi}^2_\mathrm{m} /12)[12
\hat{x}\hat{y}+4(\hat{x}+\hat{y})\hat{\phi}_\mathrm{m} +
\hat{\phi}_\mathrm{m}^2]\right\}^{-3}}~.&&\label{Fperpcappr}
\end{eqnarray}
\end{widetext}
It is easy to verify that Eq. (\ref{Fperpcappr}) reduces,
respectively, to the steady state (Eq. (\ref{expBC})) when $x=0$
and $y=0$, to the transient case in Fig. \ref{FIG5}a when $x=0$
(Eq. (\ref{expa})) and to the transient case in Fig. \ref{FIG5}d
when $y=0$ (Eq. (\ref{FperpbC})). Following the treatment of the
transient situations in Fig. \ref{FIG5}a and in Fig. \ref{FIG5}c
it is possible to plot, for this case too, a normalized
expression for the transient force, i.e. the usual $\hat{F}$, as
a function of the curvilinear position of the test particle (s=0
indicates the entrance of the magnet) for different values of
$\Delta \hat{s} = \Delta s \gamma^3/R$ and for different magnet
lengths. In Fig. \ref{FIG8}, Fig. \ref{FIG9} and Fig. \ref{FIG10}
we compared our analytical results with numerical results from
$\mathrm{TRAFIC}^4$, for the cases $\Delta \hat{s} = 0.2$,
$\Delta \hat{s} = 1.0$ and $\Delta \hat{s} = 5.0$ respectively.

\section{\label{sec:transientbunch}TRANSVERSE INTERACTION BETWEEN AN
ELECTRON AND A BUNCH ENTERING A BEND FROM A STRAIGHT PATH}

In the previous Section we dealt with all the possible
configurations for a two-particle system moving in an arc of a
circle. Now, one can consider a bunch moving on the same
trajectory and calculate the transverse force on a test particle
as the sum of contributions from all the source particles within
the bunch.

As an example, we will study here the case of a bunch entering a
long bending magnet. Such a case is important, as mentioned
before, for code benchmark purposes and for direct application in
restricted regions of parameters (negligible transverse bunch
size and bunch energy spread).

First we will analyze the case of a bunch with rectangular density
function, and we will assume the test particle to be behind the
bunch. Such an analysis will be performed using our line model
and it is therefore valid only for transverse dimension of the
bunch much smaller than the distance between the test particle
and the bunch $\Delta s_\mathrm{min}$. After the discussion, in
Section \ref{sec:transientwo}, about a two-particle system with
the test particle behind the source electron, one is led to
conclude that, within an electron bunch, interactions between
sources in front of the test particle and the test particle
itself are important and, in general, they must be responsible,
at the entrance and at the exit of the bending magnet, for sharp
changes in the transverse forces acting on the test electron. The
quantitative change depends, of course, on the position of the
test particle inside the bunch: the extreme cases are for the
test particle at the head of the bunch, where there are just
interactions with electrons behind the test particle (no head-tail
interactions), and for the test particle at the tail of the
bunch, where all the sources are in front of it (only head-tail
interactions). It may be worthwhile to underline that the sharp
jumps in the transverse force are expected to take place in a
space interval comparable, at most, with half of the bunch length,
in the case of the test electron at the tail of the bunch. In
order to show this, one can easily calculate the transverse force
acting on a test particle behind a bunch with rectangular density
distribution entering a hard-edge magnet. If, as usual,  we
indicate with $\Delta s_\mathrm{max}$ the distance from the test
particle to the head of the bunch and with $\Delta s_\mathrm{min}$
the distance from the test particle to the tail of the bunch, then
one can easily derive such an expression  from Eq.
(\ref{retcondht}) and Eq. (\ref{Fperpht}):

\begin{widetext}
\begin{equation}
{F_\mathrm{\bot~HT}^\mathrm{B}}(\phi) \simeq  \Bigg\{
\begin{array}{c}
0~~~~~~~~~~~~~~~~~~~~~~~~~~~~~~~~~~~~~~~\phi<0\\  e^2 \lambda_0
/(4 \pi \varepsilon_\mathrm{0} R) ~ \ln\left[\Delta
s_\mathrm{max}/(\Delta s_\mathrm{max} - R\phi(1+\beta))\right]
~~~~~~~~0<\phi< (\Delta s_\mathrm{max}-\Delta s_\mathrm{min})/[R(1+\beta)]\\
e^2 \lambda_0 /(4 \pi \varepsilon_\mathrm{0} R) ~ \ln[\Delta
s_\mathrm{max}/\Delta
s_\mathrm{min}]~~~~~~~~~~~~~~~~~~~~~~~~~~~~~~~~~~\phi> (\Delta
s_\mathrm{max}-\Delta s_\mathrm{min})/[R(1+\beta)]
\end{array}~, \label{HTbunch}
\end{equation}
\end{widetext}
where "HT" stands for "head-tail".

\begin{figure*}
\includegraphics*[width=180mm]{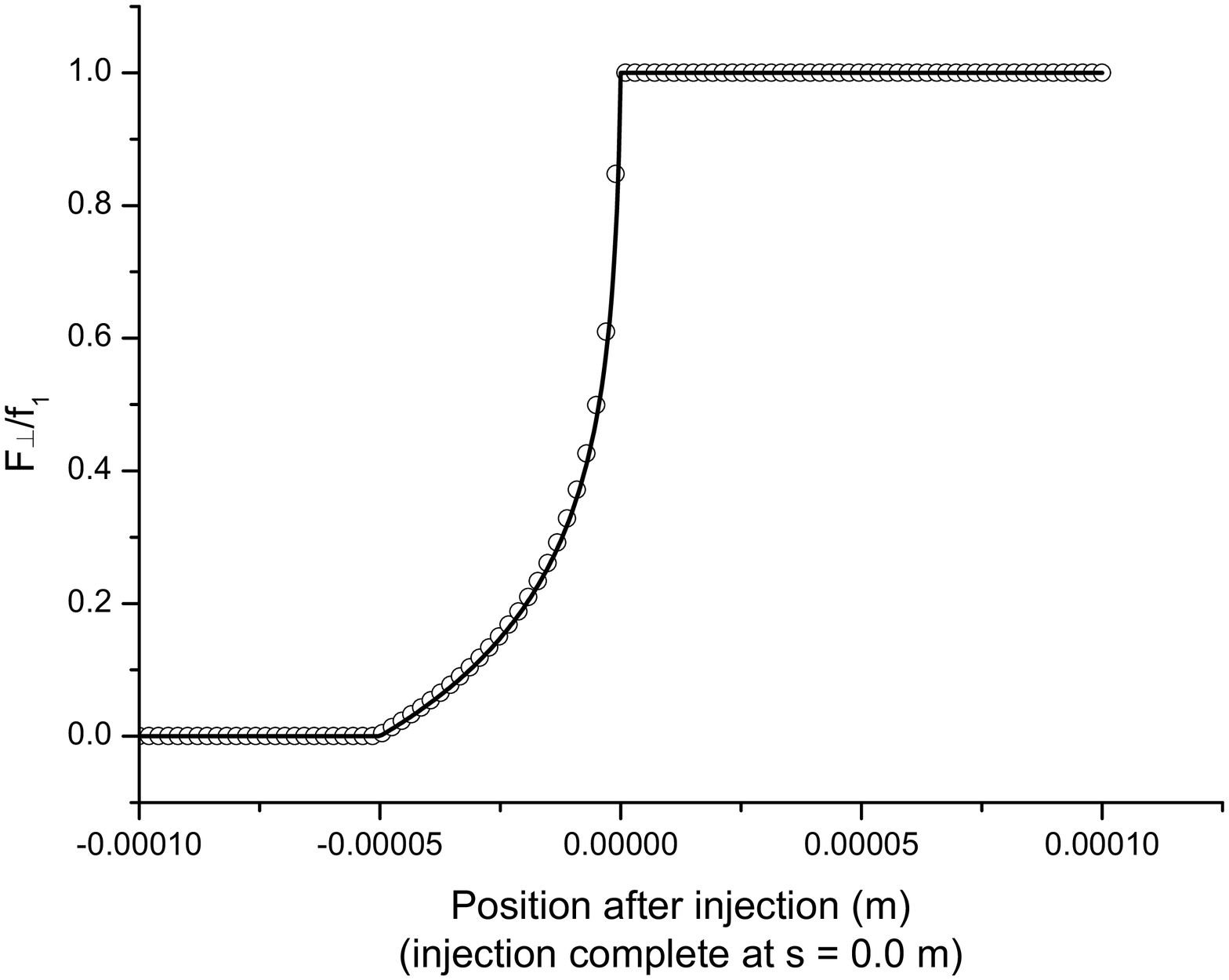}
\caption{\label{OutputHT} Normalized transverse force
$F_\mathrm{\bot}/f_1$ acting on a test particle behind a bunch
with rectangular density distribution. As the bunch enters a
hard-edge bending magnet we plot the normalized force as a
function of the position of the test particle inside the magnet
(solid line) and we compare with results by $\mathrm{TRAFIC}^4$
(circles). Parameters choice are: bunch length $ = 100 \mu m$,
$\gamma = 100$, $R=1 m$; the value of $\Delta s_\mathrm{min}$ is
$1 \mu$m.}
\end{figure*}

The following step is to actually plot the transverse force in
Eq. (\ref{HTbunch}). It is convenient to choose, as normalization
factor for the transverse force, the value $f_1 = {e^2 \lambda_0
/(4 \pi \varepsilon_\mathrm{0} R)} \ln[\Delta
s_\mathrm{max}/\Delta s_\mathrm{min}]$. Our results, compared,
once again, with simulations by $\mathrm{TRAFIC}^4$, are shown in
Fig. \ref{OutputHT} for a bunch length $\Delta
s_\mathrm{max}-\Delta s_\mathrm{min} = 100~ \mu$m, $\gamma = 100$,
$R=1$ m and for $\Delta s_\mathrm{min} = 1 ~\mu$m. This means
that, according to the validity limits of our model, this result
can be applied only when $h \ll 1 ~\mu$m. Nevertheless, this
simple example shows that the code TRAFIC$^4$ is actually able to
account for head-tail interactions. Note that, as expected, the
transient has a spatial extent of about one half of the bunch
length ($ \Delta s_\mathrm{max} / (1+\beta) $).

 We will now analyze the case of a bunch with
rectangular density function with the test particle in front of
the bunch, as depicted in Fig. \ref{FIG4}. In the injection case
we have contributions from retarded sources both in the bend and
in the straight line before the bend. The contribution from the
retarded sources in the magnet is given, basically, by Eq.
(\ref{expBtot}), and reads

\begin{equation}
{F_\mathrm{\bot m}^\mathrm{B}} \simeq  { e^2 \lambda_0 \over {4
\pi \varepsilon_\mathrm{0} R}} \left[
\ln\left({\hat{\phi}_\mathrm{max}\over{\hat{\phi}_\mathrm{min}}}\right)
+
{4\over{4+\hat{\phi}^2_\mathrm{max}}}-{4\over{4+\hat{\phi}^2_\mathrm{min}}}
\right]~, \label{shortBR2}
\end{equation}
where "m" reminds that the contributions treated by Eq.
(\ref{shortBR2}) are all from the "magnet". All that is left to do
now, is the investigation of the values which
$\hat{\phi}_\mathrm{min}$ and $\hat{\phi}_\mathrm{max}$ assume.
Let us first define with $\hat{\phi}^*$ the solution of the
retardation equation $\Delta \hat{s}_\mathrm{min} =
\hat{\phi}^*/2 + \hat{\phi}^{*3}/24$. If $\hat{\phi}^* <
\hat{\phi}$, the retarded position of the first source particle
is in the bending magnet, and $\hat{\phi}_\mathrm{min} =
\hat{\phi}^*$. On the other hand, when $\hat{\phi}^* >
\hat{\phi}$ there are no contributions to the transverse force
from the bend. Next, we define with $\hat{\phi}^{**}$ the
solution of $\Delta \hat{s}_\mathrm{max} = \hat{\phi}^{**}/2 +
\hat{\phi}^{**3}/24$. Supposing $\hat{\phi}^* < \hat{\phi}$, if
$\hat{\phi}^{**} < \hat{\phi}$ too, then all the particles
contribute from the bend, and $\hat{\phi}_\mathrm{max} =
\hat{\phi}^{**}$. On the other hand, when $\hat{\phi}^{**} >
\hat{\phi}$, we have a mixed situation, in which part of the
particles contribute from the bend and others from the straight
line before the magnet. In this case $\hat{\phi}_\mathrm{max} =
\hat{\phi}$.

The contribution from the retarded sources in the straight path
before the bend is given by

\begin{equation}
{F_\mathrm{\bot s}^\mathrm{B}} = \int_\mathrm{\Delta
\hat{s}_\mathrm{min}}^{\Delta \hat{s}_\mathrm{max}}
{R\over{\gamma^3}} F_\mathrm{\bot}(\hat{y}(\Delta \hat{s}),
\hat{\phi}) d\Delta \hat{s}~, \label{bunchcontrtrans}
\end{equation}
where "s" stands for "straight path", and where the expression for
$F_\mathrm{\bot}$ in the integrand is given by Eq. (\ref{expa}).
It is convenient, as done before, to switch the integration
variable from $\Delta \hat{s}$ to $\hat{y}$. The Jacobian of the
transformation is given by (see \cite{SALLONG})

\begin{equation}
{d\Delta \hat{s}\over{d \hat{y}}} \simeq
{(\hat{\phi}+\hat{y})^2+\hat{\phi}^4/4\over{2
(\hat{\phi}+\hat{y})^2}} \label{Jacobiantransient}
\end{equation}
After substitution of Eq. (\ref{Jacobiantransient}) and Eq.
(\ref{expa}) in Eq. (\ref{bunchcontrtrans}), one can easily carry
out the integration, thus getting

\begin{eqnarray}
{F_\mathrm{\bot \mathrm{s}}^\mathrm{B}} \simeq  {2 e^2 \lambda_0
\over {4 \pi \varepsilon_\mathrm{0} R}} \Bigg[ {\hat{\phi}\left(4
\hat{y}_\mathrm{min} + 2 \hat{\phi} + \hat{\phi}^3 \right)\over{4
\hat{y}_\mathrm{min}^2 + 8 \hat{y}_\mathrm{min}\hat{\phi} + 4
\hat{\phi}^2 + \hat{\phi}^4}}\nonumber\\&& \cr-
{\hat{\phi}\left(4 \hat{y}_\mathrm{max} + 2 \hat{\phi} +
\hat{\phi}^3 \right)\over{4 \hat{y}_\mathrm{max}^2 + 8
\hat{y}_\mathrm{max}\hat{\phi} + 4 \hat{\phi}^2 +
\hat{\phi}^4}}\Bigg]&&~. \label{bunchtransfinale}
\end{eqnarray}
As done before for $\hat{\phi}_\mathrm{min}$ and
$\hat{\phi}_\mathrm{max}$, we can now investigate the values of
$\hat{y}_\mathrm{min}$ and $\hat{y}_\mathrm{max}$. Let us start
with $\hat{y}_\mathrm{min}$. First, we define with $\hat{y}^*$ the
solution of the retardation condition $\Delta
\hat{s}_\mathrm{min} = (\hat{\phi}+\hat{y}^*)/2 +
(\hat{\phi}^{*3}/24) (4 \hat{y}^*
+\hat{\phi})/({\hat{y}^*+\hat{\phi}})$. If $\hat{y}^*>0$,  the
retarded position of the first source particle is in the straight
line before bending magnet, and $\hat{y}_\mathrm{min} =
\hat{y}^*$. On the other hand, when $\hat{y}^* < 0$,  the
retarded position of the first source particle is in the bend,
and $\hat{y}_\mathrm{min} = 0$.

Next, we define with $\hat{y}^{**}$ the solution of $\Delta
\hat{s}_\mathrm{max} = (\hat{\phi}+\hat{y}^*)/2 +
(\hat{\phi}^{*3}/24) (4 \hat{y}^*
+\hat{\phi})/({\hat{y}^*+\hat{\phi}})$. Consider the case
$\hat{y}^{**} < 0$: all the particles contribute from the bend,
that is we entered the steady-state situation. In this case
$\hat{y}_\mathrm{max} = \hat{y}_\mathrm{min}=0$. On the other
hand, when $\hat{y}^{**} > 0$, we have again a mixed situation,
in which part of the particles contribute from the bend and
others from the straight line before the magnet. In this case
$\hat{y}_\mathrm{max} = \hat{y}^{**}$.

\begin{figure*}
\includegraphics*[width=180mm]{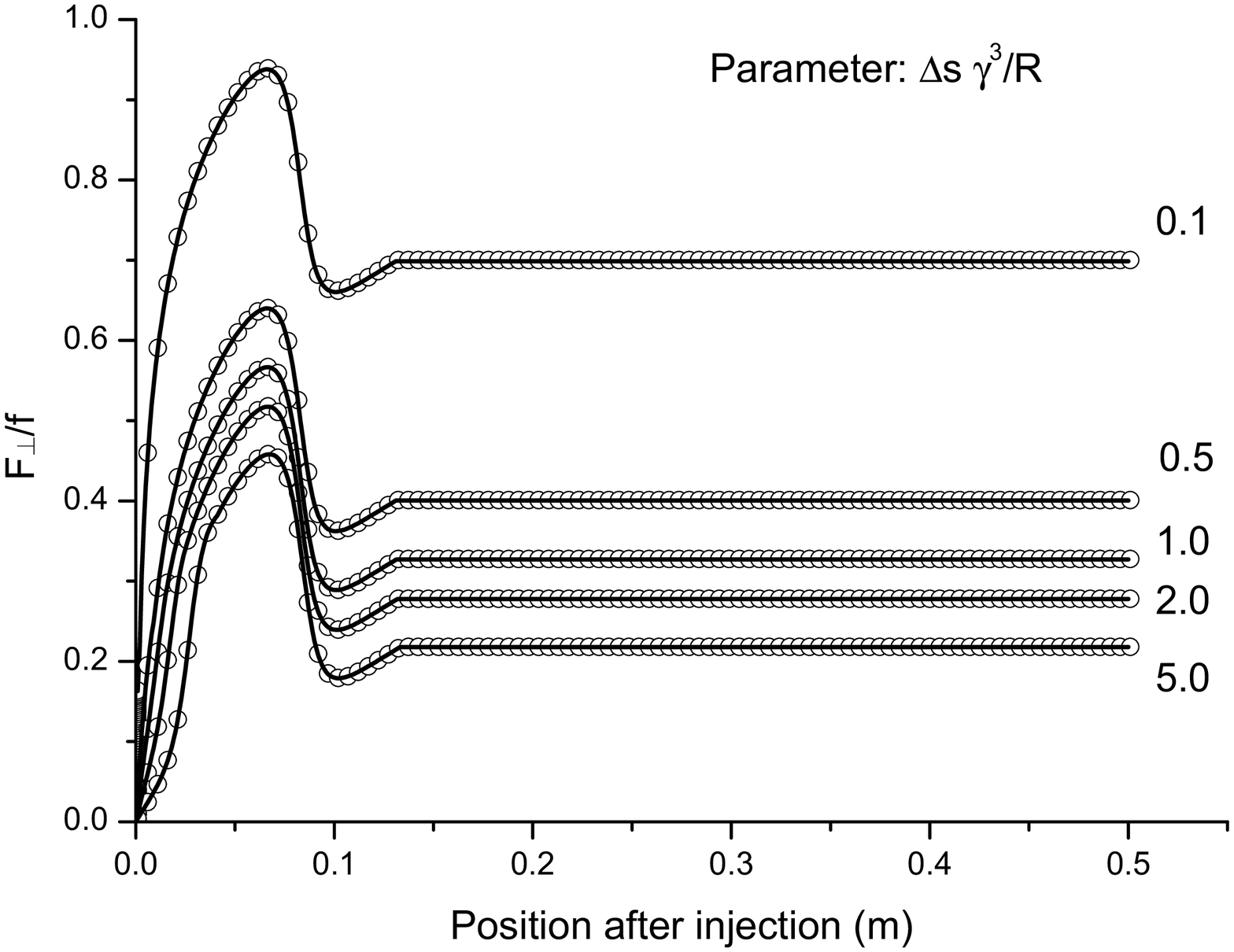}
\caption{\label{FIG11} Normalized transverse force
($F_\mathrm{\bot}/f$) acting on a test particle from a bunch with
rectangular density distribution entering a hard-edge bending
magnet as a function of the position of the test particle inside
the magnet. The solid lines show analytical results; the circles
describe the outcome from $\mathrm{TRAFIC}^4$. We chose $\Delta
s_\mathrm{max} = 100 \mu$m, $\gamma = 100$, $R=1$ m; graphs are
plotted for several values of the parameter $\Delta
\hat{s}_\mathrm{min}$.}
\end{figure*}

The following step is to actually plot the transverse force on an
electron from a bunch with rectangular distribution entering a
long bend. It is convenient to choose, as normalization factor for
the transverse force, the value $f = {e^2 \lambda_0 /(4 \pi
\varepsilon_\mathrm{0} R)} \ln(\Delta \hat{s}_\mathrm{max})$. Our
results, compared, once again, with simulations by
$\mathrm{TRAFIC}^4$, are shown in Fig. \ref{FIG11} for a bunch
length of $100~ \mu$m, $\gamma = 100$, $R=1$ m and for different
values of $\Delta \hat{s}$.

\begin{figure*}
\includegraphics*[width=180mm]{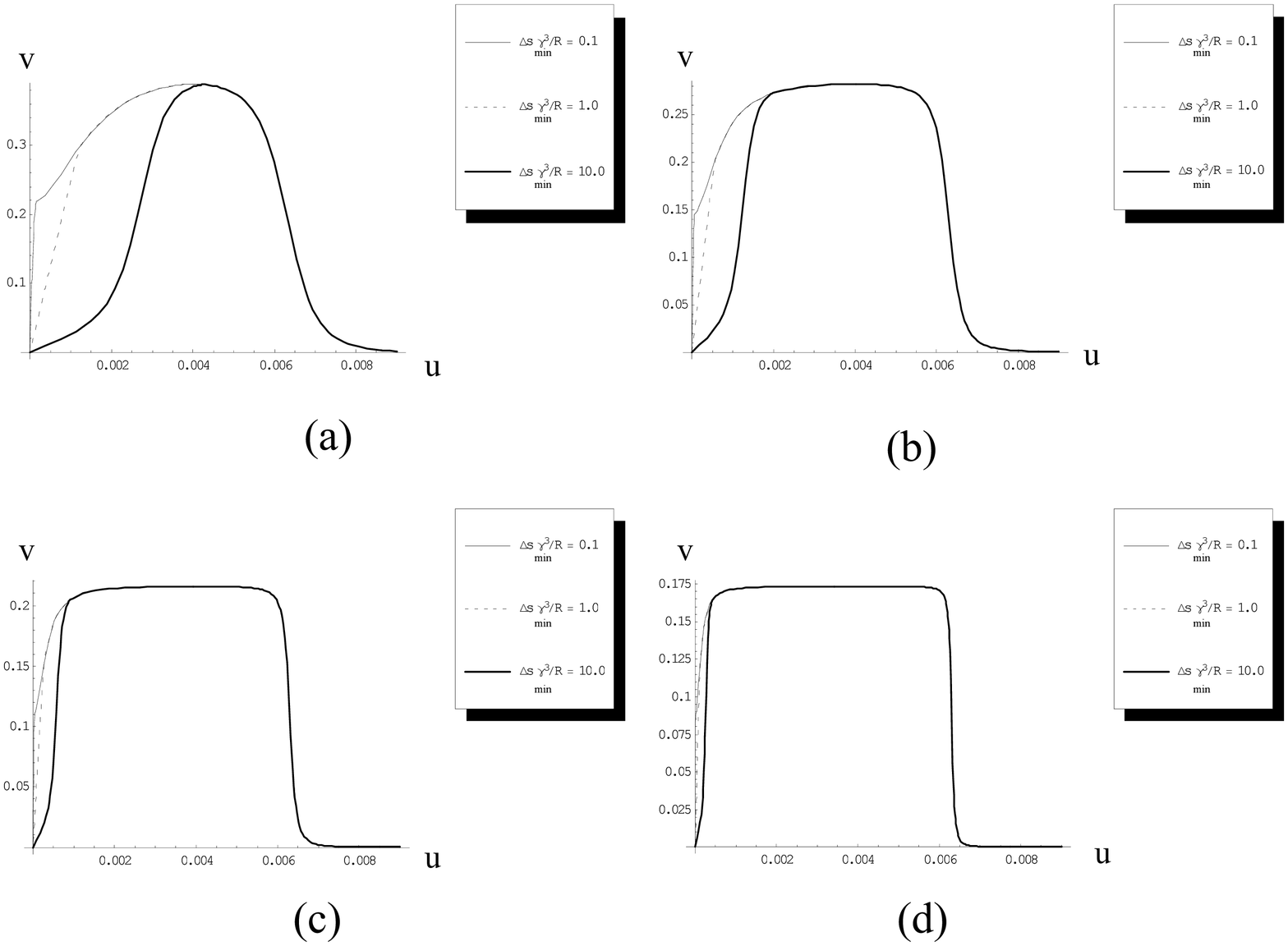}
\caption{\label{EQ60} Plot of $v=F^\mathrm{B}_\mathrm{\bot s}/f$
as a function of $u=R\hat\phi/(\gamma\hat\phi_\mathrm{b})$; here
$R=1$m and $\gamma = 100$. Case (a) the bunch length is $10^2
\mu$m, (b) $10^3 \mu$m, (c) $10^4 \mu$m and (d) $10^5 \mu$m. Each
case is parameterized with respect to different values of $\Delta
\hat{s}_\mathrm{min} = \gamma^3 \Delta s_\mathrm{min}/R$ (see the
legends in the plots). }
\end{figure*}

It is interesting to calculate the asymptotic expression for
$F^\mathrm{B}_\mathrm{\bot s}$ in the limit for a long bunch
($\Delta \hat{s}_\mathrm{max} \gg 1$) and for a short distance
between the test particle and the head of the bunch ($\Delta
\hat{s}_\mathrm{min} \ll 1$). First let us indicate with
$\hat{\phi}_\mathrm{b}$ the normalized angular extension of the
bunch ($\hat{\phi}_\mathrm{b}=\gamma \phi_\mathrm{b}$, where
$\phi_\mathrm{b}$ is just the ratio between the bunch length and
the radius of the circle $R$). By means Eq. (\ref{retcond:a}) it
is easy to prove that, when $\hat{\phi}/\hat{\phi}_\mathrm{b} <
2^{-2/3}$, $\hat{y}_\mathrm{max}$ takes bigger and  bigger values,
with an upper limit $\hat{y}_\mathrm{max} = 2 \Delta
\hat{s}_\mathrm{max}$. On the other hand, when
$\hat{\phi}/\hat{\phi}_\mathrm{b} > 2^{-2/3}$,
$\hat{y}_\mathrm{max}$ takes always decreasing values, with a
lower limit $\hat{y}_\mathrm{max} = 0$. Assuming
$\hat{y}_\mathrm{min} \ll 1$, one may check that, in the long
bunch limit, Eq. (\ref{bunchtransfinale}) reads

\begin{equation}
F^\mathrm{B}_\mathrm{\bot s} \simeq \Bigg\{
\begin{array}{c}  {e^2 \lambda_0 / (2 \pi \varepsilon_\mathrm{0}
R)}~~~~~~\hat{\phi}/\hat{\phi}_\mathrm{b} < 2^{-2/3}\\
0~~~~~~~~~~~~~~~~~~~~~~~ \hat{\phi}/\hat{\phi}_\mathrm{b} >
2^{-2/3}
\end{array}~,\label{step}
\end{equation}
which is a boxcar function. Note that in the passage from Eq.
(\ref{bunchtransfinale}) to Eq. (\ref{step}) we used the fact
that $\hat{\phi} \gg 1$. In order to visualize the limiting
process we plotted, in Fig. \ref{EQ60}, $F^\mathrm{B}_\mathrm{\bot
s}$, as it is given in Eq. (\ref{bunchtransfinale}) and
normalized to $f = {e^2 \lambda_0 /(4 \pi \varepsilon_\mathrm{0}
R)} \ln(\Delta \hat{s}_\mathrm{max})$ (i.e.
$v=F^\mathrm{B}_\mathrm{\bot s}/f$, in the plot), as a function
of the position after injection, normalized to $\hat
\phi_\mathrm{b}$ (i.e. $u=R\hat\phi/(\gamma\hat\phi_\mathrm{b})$,
in the plot). The plots in Fig. \ref{EQ60}a, b, c and d refer to
different bunch lengths (respectively $10^2 \mu$m, $10^3 \mu$m,
$10^4 \mu$m and $10^5 \mu$m). For every choice of the bunch
length we show different outcomes for several choices of $\Delta
\hat{s}_\mathrm{min}$. As one can see from Fig. \ref{EQ60}, in
the limit for $\Delta \hat{s}_\mathrm{min}\ll 1$ and $\Delta
\hat{s}_\mathrm{max} \gg 1$, one approaches the boxcar function
described by Eq. (\ref{step}). Note that, here, the radius of the
bend is $R=1$m and $\gamma = 100$, therefore we have $2^{-2/3}
R/\gamma \simeq 6.3\cdot 10^{-3}$m, while the maximum value for
$v$, due to the normalization choice, is given, from Eq.
(\ref{step}), by $2/\ln(\hat \Delta s_\mathrm{min})$. In the case
of Fig. \ref{EQ60}d, for example, we have $2/\ln( \Delta
\hat{s}_\mathrm{min}) \simeq 0.174$ (in agreement, of course, with
the maximum value found in the plot).

On the other hand, again in the same limits, one can find the
asymptotic expression for the contribution from particles with
retarded position in the bend: this is given, indeed, by Eq.
(\ref{shortBR2}), with the limit expressions for
$\hat{\phi}_\mathrm{min}$ in the case of a small distance $\Delta
\hat{s}_\mathrm{min}$, and $\hat{\phi}_\mathrm{max}$ in the long
bunch limit. For $\hat{\phi}_\mathrm{min}$ we have, from the
retardation condition in Eq. (\ref{retcond:a})
$\hat{\phi}_\mathrm{min} = 2 \hat{s}_\mathrm{min}$.

For $\hat{\phi}_\mathrm{max}$, if $\Delta
\hat{s}_\mathrm{max}<\hat{\phi}^3/24$ we have, again from Eq.
(\ref{retcond:a}), $\hat{\phi}_\mathrm{max} =
\hat{\phi}_\mathrm{b} = (24 \Delta \hat{s}_\mathrm{max})^{1/3}$.
Otherwise, when $\Delta \hat{s}_\mathrm{max}>\hat{\phi}^3/24$ then
$\hat{\phi}_\mathrm{max} = \hat{\phi}$, and we have the steady
state situation.

By means of these results one can build an expression for the
transverse force in the case of a bunch with general density
distribution $\lambda$, by considering it as a composition of
rectangular bunches with length $(s-s')$ and linear density $ds'
d\lambda(s')/ds'$, under the constraint that the important bunches
for such a composition are long enough to neglect all linear terms
in the expression for the retardation condition (i.e. $\lambda(s)
\gamma^3/R \gg  d\lambda(s')/ds'$, see \cite{SALLONG}). In this
case, the contribution from particles in the straight line reads

\begin{eqnarray}
{F_\mathrm{\bot s}^\mathrm{tot}} \simeq
\int_\mathrm{-\infty}^{s-R\phi^3/6} F_\mathrm{\bot
s}^\mathrm{B}(\phi, s-s') {d \lambda(s')\over {d s'}} d s'
\nonumber\\&&\cr = { e^2 \over {2 \pi \varepsilon_\mathrm{0} R}}
\lambda(s-R\phi^3/6) && \label{straightlambda}
\end{eqnarray}
Note that the latter expression does not depend on $\Delta
s_\mathrm{min}$. On the other hand the contribution from
particles in the bend is

\begin{eqnarray}
{F_\mathrm{\bot m}^\mathrm{tot}} \simeq
\int_\mathrm{-\infty}^{s-s_\mathrm{min}} F_\mathrm{\bot
m}^\mathrm{B}(\phi, s-s') {d \lambda(s')\over {d s'}} d s'
\nonumber\\&&\cr = { e^2 \over {2 \pi \varepsilon_\mathrm{0} R}}
\int_\mathrm{-\infty}^{s-s_\mathrm{min}} {1\over{2}} \Bigg[-1 +
{1\over{3}} \ln\left({24 \gamma^3\over{R}} {\Delta
s}_\mathrm{max}\right) \nonumber\\&&\cr- \ln\left({2
\gamma^3\over{R}} {\Delta s}_\mathrm{min}\right) \Bigg]{d
\lambda(s')\over {d s'}} ds' ~,&& \label{straightlambda2}
\end{eqnarray}
which, instead, depends on $\Delta s_\mathrm{min}$. It is very
interesting to show that this dependence on $\Delta
s_\mathrm{min}$ cancels with the dependence on $\Delta
s_\mathrm{min}$ of the steady-state force: in fact this
constitutes a general result independent from the choice of the
position of the test particle. In order to show this, let us
first note that  Eq. (\ref{straightlambda2}) can be written as
\begin{eqnarray}
{F_\mathrm{\bot m}^\mathrm{tot}} \simeq  { e^2  \over {4 \pi
\varepsilon_\mathrm{0} R}} \times \nonumber\\&&\cr \times \Bigg\{
\int_\mathrm{s-\phi^3R/24}^{s-s_\mathrm{min}}  \left[-1+
\ln(\hat{\phi}_\mathrm{b}(s')) 
-\ln(\hat{\phi}_\mathrm{min}) \right]{d \lambda(s')\over {d s'}}
ds' + \nonumber\\&&\cr + 
\int_\mathrm{-\infty}^{s-\phi^3R/24}
\left[-1+\ln(\hat{\phi})
- \ln(\hat{\phi}_\mathrm{min}) \right] {d \lambda(s')\over {d s'}}
ds'\Bigg\}&&~, \label{trans}
\end{eqnarray}
while the steady state contribution is given by
\begin{eqnarray}
{F_\mathrm{\bot steady}^\mathrm{tot}} \simeq  { e^2  \over {4 \pi
\varepsilon_\mathrm{0} R}} \times \nonumber\\&&\cr \times
\int_\mathrm{-\infty}^{s-s_\mathrm{min}} \left[-1+
\ln(\hat{\phi}_\mathrm{b}(s')) 
-\ln(\hat{\phi}_\mathrm{min})\right]{d \lambda(s')\over {d s'}}
ds'~. && \label{steady}
\end{eqnarray}
Subtracting side by side Eq. (\ref{steady}) from Eq.
(\ref{trans}), and adding the contribution from the straight
path, one finally gets the following "regularized" expression for
the transient transverse force:

\begin{eqnarray}
{\widetilde{F}_\mathrm{\bot}^\mathrm{tot}} \simeq  { e^2 \over {2
\pi \varepsilon_\mathrm{0} R}} \Bigg[\lambda(s-R\phi^3/6)-
\nonumber\\&&\cr - {1\over{6}}\int_\mathrm{-\infty}^{s-R\phi^3/24}
 \ln\left(24 (s-s')\over{R\phi^3}\right){d
\lambda(s')\over {d s'}} ds' \Bigg]~,&& \label{final}
\end{eqnarray}
which is completely independent from $\Delta s_\mathrm{min}$ and,
therefore, free from singularity in the limit $\Delta
s_\mathrm{min} \rightarrow 0$. It might be worthwhile to remark
that usual regularization techniques take place, in the study of
longitudinal (CSR) self-interactions (see \cite{SALLONG}) at the
stage of the two-particle system and \textit{before} integration
of the contributions from all the retarded sources within the
bunch. The situation is reversed here, where regularization takes
place \textit{after} integration.

\section{\label{sec:cancel}TRANSVERSE DYNAMICS AND CANCELLATION OF CENTRIFUGAL FORCES}

We will discuss here, from a qualitative viewpoint, the effect of
perpendicular forces on the transverse beam dynamics (although, as
we stated in Section \ref{sec:intro}, this article is mainly
devoted to the study of electrodynamical effects). In particular
we will concentrate on the controversial issue of the cancellation
of the centrifugal force, in the transverse equation of motion of
a particle, by the ratio between the electron energy deviation of
from its nominal value and the design radius $R$ (see \cite{RUI2},
\cite{LEE}... \cite{STUP}, \cite{LCLS}).

To illustrate this cancellation as it is explained in literature,
let us consider a bunch moving in a circular orbit with design
radius $R$, and let us indicate with $X$ the transverse
displacement of a test electron from the equilibrium orbit. Then
(see, for example, \cite{RUI2}), up to the first order in $X$,
one can easily write the equation for the transverse motion of
the test electron as:

\begin{equation}
{d^2 X \over{c^2 dt^2}}+ {X \over{R^2}} =  {\Delta E_0
\over{RE_0}}+{F_r\over{E_0} },\label{trmotion}
\end{equation}
where $E_0=\gamma_0 m c^2$ is the design energy (which is  linked
to the equilibrium radius by the relation $\beta \gamma_0 m c = e
B R$), $\Delta E_0 = (\gamma - \gamma_0)m c^2 $ is the kinetic
energy deviation, from $E_0$ and, finally, $F_r$ is the
self-interaction force in the transverse direction.

We already discussed (see Section \ref{sec:steadybunch}) the fact
that the transverse force on a test electron, both on-orbit
($X=0$) and off-orbit ($X \neq 0$) can be written, in the long
bunch limit, as the sum of two terms: a logarithmic, centrifugal
term, and a constant, centripetal one.

In \cite{LCLS},  the case of bunch compression by means of a
magnetic chicane (at LCLS) is analyzed, and the centrifugal term
is told to be essentially (aside for a negligible residual)
canceled by part of the $\Delta E_0$-term in Eq. (\ref{trmotion}).
In fact all contributions to the transverse emittance growth are
attributed to the centripetal force "which originates from
radiation of trailing particles and depends on the local charge
density along the bunch. The maximum force takes place at the
center of the bunch and its effect on the transverse emittance is
estimated in the reference. This estimate predicts an emittance
growth of $\ll 1\%$ for the worst case (last dipole of chicane-2
where the bunch is shortest)" (quoted from \cite{LCLS}).

Here we will discuss this effect, well known to the CSR community
as "the cancellation effect", within the limits of our line
model, and show that, in contrast to what has been implicitly
assumed in \cite{LCLS}, it has no general validity.

As a general remark to all the previous analysis of the problem,
we must say that, up to now, only the situation of sources behind
the test particle has been discussed, while we know, from Section
\ref{sec:transientwo}, that head-tail interactions are present
too: they are characterized, at least within the limits of our
model, by a magnitude of the same order of the tail-head
interactions and the spatial extension of their transient is
negligible (comparable, at most, with half of the bunch length).
Since this centrifugal force depends on the position of the test
particle along the bunch, it will be responsible for normalized
emittance growth.

Anyway, let us consider, in particular, the tail-head interaction
problem. As underlined in Section \ref{sec:intro}, the full
solution to the evolution problem is met when one is able to
solve simultaneously the equation of motion and the equation for
the electromagnetic field. Then, in principle, we may adopt, in
our discussion, two separate viewpoints: in the first one we
imagine to solve, in some way, the full evolution problem, while
in the second we treat the same situation by means of a
perturbation theory approach, assuming that, in first
approximation, the motion of a rigid bunch is driven by the
external magnetic field alone and then calculating the
perturbation to the particle motion due to transverse
self-fields: we will show that, in both cases, the cancellation
effect has no general validity.

Let us begin with the first viewpoint. Generally speaking, if
there was \textit{complete} compensation between the
\textit{total} transverse force and the $\Delta E_0$-term, then
the particles with the same total energy (assumed equal to the
sum of kinetic energy and potential self-energy from the bunch)
would have followed the same trajectory. Such a complete
cancellation would be based upon two assumptions. Firstly,
$\Delta E_0 = - e \Delta \Phi$ and, secondly, $F_{\bot} = e
\Phi/R$. Note that, in this case, $\Delta E_0 + R F_{\bot} = e
\Phi(0)$. This, of course, is not the case, since the
cancellation has always been understood for the centrifugal term
alone (and, anyway, it is not complete). Moreover let us remind
that, as explained in Section \ref{sec:steadybunch}, there is no
physical ground to distinguish between the centripetal term and
the centrifugal one: from a physical viewpoint there is just an
overall centrifugal force. A further subdivision is just of
mathematical nature, and explains how the transverse force
plotted in Fig. \ref{FIG2} behaves from one asymptotic (short
distance between test and source particle) to the other (large
distance between test and source particle). This fact, alone,
suggests that the cancellation issue has no general validity and
that, indeed, it is artificial (as the subdivision between
centripetal and centrifugal term is) even when it works, like in
the case of a coasting beam in a simple, circular steady-state
motion.

Nevertheless, let us retain, in our terminology, the distinction
between centripetal and centrifugal term. Then, a
\textit{complete} cancellation between \textit{centrifugal} force
and potential term would be  based, similarly as before, upon two
facts. Firstly, $\Delta E_0 = -e \Delta  \Phi$ and, secondly,
$F_{\bot}^\mathrm{centrifugal} = e \Phi/R$, at any time. Whenever
one of these assumptions is no more verified, the cancellation
fails to happen.

Of course, in order to show that the cancellation effect is not
valid in general, it is sufficient to provide a counter-example.
Let us consider, therefore, the case of a bunch with zero initial
energy spread or, more simply, a two-particle system with the test
electron in front of the source, such that $\Delta \gamma =
\gamma - \gamma_0 = 0$. Let us restrict to a case in which only
space charge effects are important, as concerns the longitudinal
motion: therefore we consider, in this sense, two electrons
running on a straight path.

The Coulomb (space-charge) interaction changes both the kinetic
energy of the particles and their relative distance, which we will
assume, initially, equal to $\Delta s$ in the laboratory frame.
Let us deal with the problem in the (instantaneous) rest frame of
the center of mass of the system (simply designated as the rest
frame, in the following) and assume that the particles do not
move relativistically in such a frame. After a long time one will
be left, asymptotically, with two particles far away from each
other, each one with a kinetic energy equal to $e^2/(8 \pi
\epsilon_0 \Delta \widetilde{s})$, where $\Delta \widetilde{s} =
\gamma \Delta s$. Therefore, the change in kinetic energy of the
test particle in the laboratory frame, can be found by means of a
Lorentz transformation (performed on the momentum in the
longitudinal direction):

\begin{equation}
\Delta E_0 = \gamma {c\over{\beta}}\left({2m e^2\over{8 \pi
\epsilon_0 \gamma\Delta s}}\right)^{1/2} = {c\over{\beta}}
\left({\gamma m e^2 \over {4 \pi \epsilon_0 \Delta
s}}\right)^{1/2}~. \label{ultima}
\end{equation}
On the other hand, again in the asymptotic limit, the potential
energy of the test electron (as well as the one of the source)
will approach zero. Then, at least after a long time from the
beginning of the evolution, one has

\begin{equation}
\Delta E_0(t\rightarrow \infty)  \neq -e \Delta \Phi(t\rightarrow
\infty) = e \Phi(0) ~.\label{dopoultima1}
\end{equation}
This means that the first assumption for the cancellation ($\Delta
E_0 = - e \Delta \Phi$ at every time) is violated, due to the
inclusion of space charge forces in the longitudinal motion; as a
result we can say that there is no cancellation effect.

Of course, one could also treat the situation of a circular
motion, which is more complicated, but the example we gave is
sufficient to show that the  effect is not valid in general.

\begin{figure}[htb]
\includegraphics*[width=90mm]{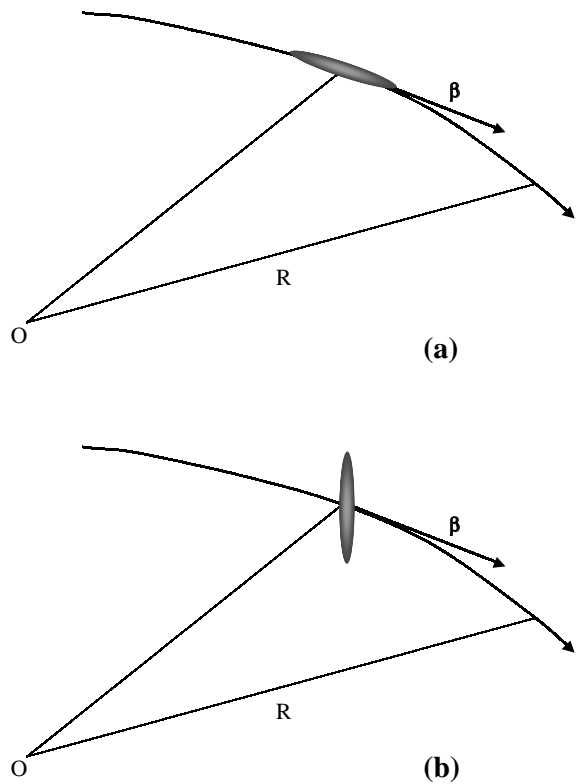}
\caption{\label{FIG12} Schematic of a line bunch in a circular
trajectory; (a) the line bunch is oriented along the velocity; (b)
the line bunch is oriented perpendicularly with respect to the
velocity.}
\end{figure}

We are now left to deal with the second viewpoint. As it has
already been said above, by following this approach, we assume
that, in zero approximation, the particles move under the action
of the external field alone and then, once the particles motion
is fixed (in the zeroth order), one can calculate the
perturbation to such a motion due to self-fields. Note that, if
one treats in this framework the example of a coasting beam or of
a rigid line bunch with finite longitudinal dimension, then $\Phi$
is constant, which suggest there are no more contradictions with
the cancellation assumptions. Nevertheless we will now show that
cancellation is far from being a general effect. Let us study,
first, the case of a rigid bunch. Consider, for example, the two
cases of a rigid line bunch oriented either along the velocity or
perpendicular to it (see Fig. \ref{FIG12}).

For simplicity we can study the non-relativistic case: in the
relativistic one we have an analogous situation. The kinetic
energy of the line bunch and the scalar potential acting on each
particle do not depend on the orientation of the bunch with
respect to the velocity vector. If the cancellation effect had
general validity, also the transverse self-force should not
depend on the orientation but, indeed, one can easily verify that,
for the case of the line bunch oriented along the perpendicular
direction (see Fig. \ref{FIG12}b), the self-interaction force is
two times smaller than in the case of orientation along the
longitudinal direction. In fact, in the case of the bunch
oriented in the longitudinal direction we have repelling forces
parallel to the velocity vector while, in the other case, they
are perpendicular: according to Eq. (\ref{stress}),  the momentum
flux along the parallel direction (that is $T_{33}$) is then
equal to the potential energy but having opposite sign in the
first case, and equal to zero in the second. This fact shows that
the cancellation effect has no general validity. This can be
discussed in a somewhat deeper way too. As we showed in Section
\ref{sec:steadytwo}, in order to get a correct dynamical
interpretation of the electrodynamical transverse forces for a
two-particle system, one needs to account for the
self-interaction energy and momentum flux and, by doing so, the
energy and momentum of the system do not form a four-vector
anymore (as, instead, people always assume in considering the
cancellation issue). They are, indeed, components of a second
rank tensor (the energy-momentum tensor) and, therefore,
transform in a different way with respect to a four-vector: this
is the conceptual reason which explains why the cancellation is
not a fundamental effect; as soon as one cannot consider energy
and momentum as a four-vector anymore the cancellation is
spoiled, in general, by the $T_{33}$ component of the stress
tensor in the rest frame.

Also the analysis of the transient behavior leads to the
conclusion that the compensation is not of general importance.
Let us consider the case of a bunch entering a bend (which we
discussed in Section \ref{sec:transientbunch}, in the laboratory
frame. The formation length for both the scalar potential and the
transverse force (after which the bunch reaches the steady state
situation) is simply the overtaking length $2(3 s R^2)^{1/3}$.
Now, on the one hand the transverse force is zero at the
beginning of the bend, and relaxes to a $\gamma$-independent
value in the steady-state situation while, on the other hand, the
scalar potential starts from a $\gamma$-dependent value before
the bend and relaxes to a $\gamma$-independent value in the
steady state regime too. Comparing the two transients for the
scalar potential and for the transverse force,  we can therefore
conclude that the first one is characterized by an additional
free parameter (the initial value of the beam energy) with
respect to the second, and this means that the two transients are
completely independent. As a result, again, the compensation
proves to fail in this case (since it should follow from the
subtraction of the centrifugal force with the term, in $\Delta
E_0/(R E_0)$, proportional to $e\Phi/R$). Of course, again, from
a more general viewpoint, the failure of the cancellation can be
just seen as a consequence of the fact that energy and momentum
are components of a tensor, and not of a vector, that is as a
consequence of the different geometrical nature of a vector with
respect to a tensor.

As a last example, one can study, again in the perturbative
framework, the case of a beam with an initial kinetic energy
chirp (which encompasses the bunch compression case) or,
equivalently, the case of two electrons moving on rectilinear
trajectory in such a way that that, for the test particle,

\begin{equation}
\Delta T_{\mathrm{chirp}} \equiv \Delta \gamma(0) m c^2  \gg
{c\over{\beta}} \left({\gamma m e^2\over{4 \pi \epsilon_0 \Delta
s}}\right)^{1/2}~, \label{condiz}
\end{equation}
which justifies the perturbative approach.

It is possible to demonstrate that, in first order in the
self-fields, space charge induces a change in kinetic energy, in
the laboratory frame, given by

\begin{equation}
\Delta E_0 = {c\over{\beta}} \left({\gamma m e^2\over{4 \pi
\epsilon_0 \Delta s}}\right)^{1/2} { \left({\Delta
T_\mathrm{chirp} e^2\over{8 \pi \epsilon_0 \Delta
s}}\right)^{1/2}/{\Delta T_\mathrm{chirp}}} ~,\label{DTS}
\end{equation}
where we started from the rest frame, we used a first order
expansion in the kinetic energy difference, $\Delta
T_\mathrm{chirp}$, and, finally, a Lorentz transformation to the
laboratory frame. Eq. (\ref{DTS}) shows that $\Delta E_0$
includes a free parameter (equal to ${ \left({\Delta
T_\mathrm{chirp} e^2/{8 \pi \epsilon_0 \Delta
s}}\right)^{1/2}/{\Delta T_\mathrm{chirp}}}$) which depends on
the energy chirp. This makes the assumption $\Delta E_0 = -e
\Delta \Phi$ incorrect and demonstrates that the cancellation
issue does not exist at all during the bunching process.

To sum up, the final result of this investigation is that the
cancellation effect has by no means general validity and that one
must exercise extreme care when dealing with transverse dynamics
issues. For example,  the qualitative discussion in this Section
shows that, in contrast with \cite{LCLS}, when one considers the
problem of calculating the emittance dilution in a bunch
compressor, not only centripetal force but also centrifugal
contributions  must be taken into account. We can conclude that,
in practice, the total force one has to account for is, at least,
one order of magnitude larger than the so called centripetal
force.

\section{\label{sec:concl}SUMMARY AND CONCLUSIONS}

In this paper we presented a fully electrodynamical study of
transverse self-forces within an electron bunch moving in an arc
of a circle. Our analysis is based on a line-bunch model. In the
case of test particles in front of the source, our model is valid
whenever $h \ll R/\gamma^2$. On the other hand, when the source
electron is in front of the test electron, then the situation is
more complicated and the model can be applied for $|\Delta| s \gg
h$ The cases that do not fulfill these conditions are left for
future study.

In Section \ref{sec:steadytwo} we studied, first, the situation
of a two-particle system moving on a circular path, finding an
approximated expression for the transverse self-forces which is
the product of a factor dependent on the parameters that specify
the system setup and a universal function (i.e. a function
independent of such parameters): for every distance between the
particles we found a centrifugal force, which has a qualitative
and quantitative explanation by means of simple arguments from
relativistic dynamics. We concluded that both the tail-head and
head-tail interactions are important: in the first case the
Coulomb interaction plays a role besides the Radiative one, while
in the second only Radiative contributions are present.

Further on, in Section \ref{sec:steadybunch}, after discussing
the applicability region of our model, we integrated the results
for a two-particle system, thus finding an expression for the
transverse interaction between a line bunch and a particle in
front of it. Such an expression is structured as the sum of a
centrifugal logarithmic part and a centripetal term, of which we
studied the asymptotic behaviors in the limit of short and long
bunch, both with a small distance between the test particle and
the bunch head.

In particular we found that, in the limit of a short bunch, the
centripetal force tends to zero as $(\gamma
\phi_\mathrm{max})^2$, while in the limit of a long bunch we
found results already well known in literature. We were able to
explain the constant centripetal term in the latter case as an
overall effect of the transient between the asymptotic behavior
(identical in $\phi$, different in $\Delta s$) of the transverse
force in the two-particle system, respectively for small or long
distance between the two electrons. We concluded that the
centripetal term is therefore only of mathematical nature, and
there is no physical ground to distinguish it from the
centrifugal term: from a physical viewpoint there is, in fact,
just an overall centrifugal force.

In Section \ref{sec:transientwo}, again within the region of
applicability of our model, we extended our considerations to a
two-particle system moving in an arc of a circle, thus finding,
for the first time, exact and approximated analytical expressions
for the transverse force in all the possible transient
configurations (see Fig. \ref{FIG5}), including the case in which
the source particle is in front of the test particle. Furthermore
we plotted the expression for the transverse force in several
practical cases, which are important for a quick evaluation of the
magnitude of the effect and for cross-checks between computer
codes. In particular we report a very good agreement with
TRAFIC$^4$, which demonstrates that such a code can easily deal,
from a numerical viewpoint with the transverse transient problem.

In Section \ref{sec:transientbunch}, we analyzed the situation of
the transverse interaction between a line bunch and a test
particle moving in an arc of a circle. We treated, in particular
the case of the injection from a straight section into a
hard-edge bending magnet. Firstly we calculated exact and
approximated expressions for the transverse force; secondly,
following what we did in Section \ref{sec:transientwo}, we
provided a few graphical examples; thirdly we analyzed our
expressions in the limit for a long bunch and a short minimal
distance between the test particle and the head of the bunch.

We showed that the contribution from the particles whose retarded
positions are in the straight line before the bend is well
described, as a function of the normalized angular position of
the test particle inside the bend, by a boxcar function with
length equal to the normalized bunch angular length multiplied by
a characteristic constant which depends on the structure of the
retardation condition. By simple composition of rectangular
bunches we  provided an expression for the calculation of the
transverse interaction in the case of a long bunch with an
arbitrary density distribution. We showed that, in the chosen
limits, the contribution from the particles in the bend is
independent from $\Delta s_\mathrm{min}$ and that, in contrast to
this, the contribution from the particles in the straight line is
dependent on $\Delta s_\mathrm{min}$. Finally we proved that such
a dependence can be removed by subtraction of the steady-state
transverse self-interaction, thus providing a "regularized"
expression for $F_\mathrm{\bot}$.

The case of injection provides a useful example for a quick
evaluation of the transverse self-fields magnitude as well as for
computer codes benchmark. By means of the same approach one can
analyze also the case of ejection, which is left to future work.

Finally, in Section \ref{sec:cancel} we proved that the partial
compensation of transverse self-force
has by no means general validity, and one must exercise extreme
care when dealing with transverse dynamic issues. In fact, in
contrast with \cite{LCLS}, we concluded that, when one considers
the problem of calculating the emittance dilution in a bunch
compressor, not only centripetal force but also centrifugal
contributions  must be taken into account. Therefore, in practice,
the total force one has to account for is, at least, one order of
magnitude larger than the so called centripetal force.

\section{\label{sec:acknowl}ACKNOWLEDGEMENTS}

The authors wish to thank Reinhard Brinkmann, Yaroslav Derbenev,
Klaus Floettmann, Vladimir Goloviznin, Georg Hoffstaetter, Rui Li,
Torsten Limberg, Helmut Mais, Philippe Piot, Joerg Rossbach, Theo
Schep and Frank Stulle for their interest in this work.

\end{document}